\documentclass[lettersize,journal]{IEEEtran}

\usepackage{amsmath,amsfonts}
\usepackage[ruled,linesnumbered]{algorithm2e}
\usepackage[caption=false,font=footnotesize,labelfont=rm,textfont=rm]{subfig}
\usepackage{makecell}
\usepackage{stfloats}
\usepackage{ragged2e}
\usepackage{url}
\usepackage{verbatim}
\usepackage{graphicx}
\usepackage{balance}
\usepackage{cite}
\usepackage{multirow}
\usepackage{enumitem}
\usepackage{bm}
\usepackage{subfig}
\usepackage{tikz}
\usetikzlibrary{shapes, arrows.meta, positioning}
\usepackage{booktabs}
\usepackage{lipsum}
\usepackage{amssymb}
\usepackage[colorlinks,linkcolor=blue,anchorcolor=blue,citecolor=blue]{hyperref}

\usepackage{etoolbox}
\makeatletter
\patchcmd{\@makecaption}
{\\}
{.\ }
{}
{}

\newcommand{\myrefeq}[1]{(\ref{#1})}
\newcommand{\myreffig}[1]{Fig. \ref{#1}}
\newcommand{\myreftable}[1]{Table \ref{#1}}

\title{Intelligent Beamforming and Handover via Physics-Informed Beam-Aware CKM Diffusion}

\author{
	Le Zhao, Zesong Fei, \textit{Senior Member, IEEE}, Xinyi Wang, \textit{Member, IEEE}, Nan Ha, \\ Yining Wang, Yong Zeng, \textit{Fellow, IEEE}, and Rui Zhang, \textit{Fellow, IEEE}
	\thanks{
		Le Zhao, Zesong Fei, Xinyi Wang, Nan Ha, and Yining Wang are with the School of Information and Electronics, Beijing Institute of Technology, Beijing 100081, China.
		Yong Zeng is with the National Mobile Communication Research Laboratory, School of Information Science and Engineering, Southeast University, Nanjing 211189, China. Yong Zeng is also with Purple Mountain Laboratories, Nanjing 211111, China. Rui Zhang is with the Department of Electrical and Computer Engineering, National University of Singapore, Singapore 117583.
		(e-mail: tobin$\_$bit@icloud.com, feizesong@bit.edu.cn, bit$\_$wangxy@163.com, hanan@bit.edu.cn, wang$\_$yining325@163.com, yong$\_$zeng@seu.edu.cn, elezhang@nus.edu.sg)

		Part of this paper \cite{zhao2026beamckmdiff} was accepted by the 2026 IEEE International Conference on Computer Communications (INFOCOM), Tokyo, Japan.
	}
	\vspace*{-0.35cm}
}

\begin{document}
	\maketitle
	\begin{abstract}
		Downlink communications from base stations (BSs) to unmanned aerial vehicles (UAVs) in sixth-generation (6G) networks require precise beam alignment to overcome severe mobile communication path loss. However, traditional exhaustive beam sweeping relies on discrete codebooks and consumes valuable air-interface resources for online measurements, rendering it inefficient for highly dynamic aerial environments. In this paper, we propose BeamCKMDiff, a physics-informed generative diffusion framework designed to construct high-fidelity continuous beam-aware channel knowledge maps (BeamCKMs). Unlike existing empirical approaches, BeamCKMDiff employs a diffusion transformer (DiT) architecture featuring a dual-path conditioning mechanism. It integrates an analytical line-of-sight (LoS) beam prior with environmental topologies, while simultaneously embedding continuous beamforming vectors through an adaptive layer normalization (adaLN) mechanism. Building upon this differentiable generative mapping, we propose a CKM-based end-to-end continuous beamforming and proactive dual-BS handover algorithm. By analytically propagating gradients through the reverse diffusion process, the proposed framework enables successive convex approximation (SCA)-based beamforming optimization in the computational domain, effectively bypassing physical pilot scanning.
		Simulation results demonstrate that BeamCKMDiff achieves a normalized mean square error (NMSE) of --23.96 dB, establishing a highly reliable spatial prior. Compared to discrete beam sweeping and LoS-assumed baselines, the proposed CKM-based beamforming adapts to non-LoS (NLoS) conditions, providing accurate beam alignment and higher spectral efficiency. Furthermore, the integrated dual-BS handover mechanism eliminates blockage-induced link outages, ensuring robust connectivity for high-mobility aerial platforms.
	\end{abstract}


	\section{Introduction}

	\IEEEPARstart{T}{he} advancement of unmanned aerial vehicles (UAVs) has driven the fast development of low-altitude wireless networks (LAWNs) as a fundamental 6G connectivity paradigm, providing a versatile and economical solution for the expansion of non-terrestrial networks \cite{ZengYong_2021_CKM_ST}. Within this architecture, UAVs can serve as aerial users, flexibly accessing the network and leveraging their high mobility to enable dynamic task scheduling and fast topology response, thereby supporting heterogeneous network coordination \cite{commag_fei}. To satisfy the stringent capacity and reliability demands of such emerging aerial applications, massive multiple-input multiple-output (MIMO) and millimeter-wave (mmWave) technologies have been widely deployed \cite{MIMOuavCM2021}. Within these architectures, beamforming emerges as an indispensable technique that substantially enhances link performance by precisely concentrating signal energy towards desired directions to overcome severe spatial pathloss.

	{Traditional beamforming designs include linear precoders such as minimum mean-square error (MMSE) and zero forcing (ZF), together with optimization-based methods such as weighted MMSE (WMMSE) and semidefinite relaxation (SDR) for continuous precoder design} \cite{Beamforming_Mag, shi2011wmmse, luo2010sdr}.
	However, these optimization-based approaches require perfect, instantaneous channel state information (CSI), which is practically hard to obtain in highly dynamic aerial environments due to the rapid channel fluctuations and severe pilot overhead. As a practical alternative, implicit channel acquisition via beam training is regarded as an irreplaceable prerequisite \cite{wang2009beam, BeamSweeping_CM, xiao2016hierarchical, ning2023wide}. In the IEEE 802.11ad and third generation partnership project (3GPP) standards \cite{wang2009beam, BeamSweeping_CM, 3gpp_ts_38_331}, the sector-level sweep, beam refinement protocol, and beam management mechanisms have been formally established. In \cite{xiao2016hierarchical}, a joint sub-array and de-activation hierarchical codebook based on weighted summation was proposed, and \cite{ning2023wide} designed an enhanced hierarchical codebook with low sidelobe to ensure training accuracy. While traditional methods have ingeniously designed hierarchical codebooks and state-beamforming mechanisms to improve alignment accuracy, the realization of reliable UAV communication is fundamentally hindered by the high mobility of aerial platforms and the resulting rapid fluctuations in CSI. Consequently, the overhead required for real-time exhaustive beam sweeping becomes prohibitive  \cite{ref_mmwave_uav}. Conventional pilot-based channel estimation schemes not only consume significant spectral resources but also introduce latency that frequently leads to catastrophic beam misalignment and link failure \cite{YuanMissBeamwcl2020}. To alleviate this beam management burden and break the CSI acquisition bottleneck, recent studies have begun incorporating explicit side information to reduce the searching space. For instance, mapping location coordinates to coarse beam directions in sequential beam prediction \cite{va2019online, klautau20185g}, as well as multi-modal sensory data, such as LiDAR or sub-6GHz anchor signals, have been exploited to predict line-of-sight (LoS) paths and blockages \cite{klautau20185g}, effectively reducing the candidate beam searching space and underscoring the urgent necessity for environment-aware predictive communication paradigms.
	
	% le2025IMNet	ZengYong_2024_IEEE_ST

	Building upon the exploitation of such environmental priors, the channel knowledge map (CKM) has been proposed as a transformative enabler for 6G networks \cite{Romero_SPM_RadioMap, ChannelDataTWC_ZY, ren2026channel}. By leveraging the site-specific nature of radio propagation, CKM constructs a multi-dimensional database that maps spatial coordinates directly to key channel properties, thereby providing low-cost spatial prior information. While early CKM construction methods relied on computationally prohibitive ray-tracing \cite{Rizk_1997_TVT} or statistical interpolations, e.g., Kriging that struggles to capture intricate 3D urban shadowing \cite{Krige_2017_TVT}, deep learning (DL) has recently revolutionized this field.{Discriminative architectures, including convolutional neural networks \cite{levie2021radiounet, Junting_RM_CNN} and graph attention networks \cite{li2023graph}, learn deterministic radio maps, whereas generative architectures such as generative adversarial networks \cite{Zhang_RMEGAN_2023} model the distribution of spatial propagation characteristics.} More recently, denoising diffusion probabilistic models (DDPMs) have demonstrated strong CKM-generation performance. Frameworks such as CKMDiff \cite{ZengYong_CKMDiff_arxiv,RadioDiff_TCCN} apply DDPMs to recover high-frequency details and sharp shadow boundaries that regression-based methods tend to oversmooth.

	Despite the success of deep learning in CKM construction, most frameworks output scalar channel gain values tailored for single-antenna systems, failing to support the coherent phase requirements of multi-antenna beamforming. Initial explorations into beam-aware CKMs (BeamCKMs) for beamforming optimization, such as channel angle maps and beam index maps \cite{wu2023environment}, successfully reduced beam training costs, yet their explicit construction processes were largely overlooked. To address this, subsequent works introduced dedicated datasets like CKMImageNet \cite{ZengYong_tcom_ckmimagenet} and problem-specific neural architectures \cite{yang2025radio, chatelier2025model} to learn the location-to-CSI mapping; nevertheless, their reliance on computationally prohibitive ray-tracing or dense environmental sampling severely limits real-world scalability.
	As a step further, to resolve phase ambiguity, the transformer-UNet (TransUNet) based BeamCKM framework was proposed to construct CKM under specific beamforming vectors \cite{wang2025beamckm}. Parallel to this, recent efforts have introduced physics-informed BeamCKM to decouple deterministic array radiation from neural network predictions \cite{li2026u6g}, and utilized tensor decomposition to reconstruct MIMO beam coverage from sparse measurements \cite{sun2026mimo}.
	Even with these advancements, prior works fundamentally rely on measurement samples gathered under predefined, static beamforming codebooks. In highly dynamic UAV deployments, acquiring such beam-specific empirical data is practically infeasible. Moreover, they treat the beam configuration either as a scalar parameter or tie it strictly to a discrete codebook, rather than integrating the continuous beamforming vector as an explicit spatial prior. Consequently, these architectures cannot freely optimize beams based on the CKM, lacking the capability to support end-to-end beam optimization for specific locations.

	To bridge these gaps, this paper proposes BeamCKMDiff, a physics-informed generative diffusion framework designed to construct BeamCKMs for continuous beamforming optimization. By integrating a diffusion transformer (DiT) with beam-aware feature fusion, we establish a generative mapping from continuous beamforming vectors to spatial channel gains.{To the best of our knowledge, BeamCKMDiff is the first CKM framework to condition map generation on continuous beamforming vectors and to exploit the differentiability of this mapping for pilot-free beam optimization. Existing methods are restricted to fixed beams or discrete codebooks.} Building upon this, we investigate an end-to-end beamforming and handover mechanism to maintain robust UAV connectivity. The main contributions are summarized as follows:

	\begin{itemize}
		\item We propose a physics-informed BeamCKMDiff architecture to construct BeamCKMs conditioned on continuous beamforming vectors. This framework utilizes a dual-path conditioning mechanism to couple complex-valued beamforming vectors with spatial propagation. Specifically, analytical LoS beam maps are integrated with environmental topologies to provide explicit physics-based priors, while continuous beamforming vectors are embedded into the DiT backbone via an adaptive layer normalization (adaLN) mechanism.

		\item We develop a differentiable CKM-triggered beamforming design strategy based on the unrolled reverse diffusion process. This strategy utilizes the differentiability of the generative mapping to compute analytical gradients for beamforming optimization. The implementation involves executing successive convex approximation (SCA) in the computational domain to determine continuous beamforming vectors without online measurements.

		\item{We design a proactive dual-BS handover strategy that adapts to spatial channel variations. BeamCKMDiff evaluates candidate links virtually, and the radio access network (RAN) central controller pre-configures the target beam before association transfer. This decision-level coordination avoids air-interface pilot scanning and prevents blockage-induced outages.}
	\end{itemize}

	Extensive simulations validate that BeamCKMDiff achieves high construction accuracy by precisely reconstructing spatial propagation characteristics. Compared to LoS-assumed beam alignment \cite{Beamforming_Mag} schemes and discrete NR beam sweeping \cite{BeamSweeping_CM}, the proposed framework achieves higher beam alignment accuracy and link performance. Furthermore, the CKM-based{proactive} dual-BS handover effectively eliminates link outages in complex non-LoS (NLoS) scenarios, ensuring robust connectivity for high-mobility aerial platforms.

	The remainder of this paper is organized as follows. Section II introduces the aerial system model and the beam-aware channel gain characterization. Section III details the proposed BeamCKMDiff architecture and the physics-informed training strategy. Section IV discusses the proposed CKM-based beamforming and handover algorithm. Numerical results and performance analysis are presented in Section V, followed by concluding remarks in Section VI.

	\section{System Model and Problem Formulation}
	\label{sec:system_model}

	\subsection{System Model}
	\label{sec:system_model_a}
	\subsubsection{Environment and UAV model}
	We consider a downlink communication system operating in an urban environment, where a set of terrestrial base stations (BSs), denoted by $\mathcal{B}$, {establish communication links with} a high-mobility UAV. The mission area is modeled as a 3D space with dimensions $\{W_1, W_2, H\} \in \mathbb{N}^+$. To facilitate environment-aware predictive communication, the horizontal space is discretized into a uniform grid{$\tilde{\mathcal{A}}$} with spatial resolution $\Delta \kappa$. Thus, the grid dimensions are $N_1 = W_1/\Delta \kappa$ and $N_2 = W_2/\Delta \kappa$. The physical environment topology is defined by a matrix $\mathbf{E} \in \mathbb{R}^{N_1 \times N_2}$, where each element $[\mathbf{E}]_{i_x,j_y}$ represents the building height at the discrete grid location $(i_x\Delta \kappa, j_y\Delta \kappa)$.{Given the UAV flight altitude $z_{\rm h}$, building heights exceeding $z_{\rm h}$ are truncated via $\min([\mathbf{E}]_{i_x,j_y}, z_{\rm h})$ and normalized by $z_{\rm h}$ into $[0,1]$, so that the input features reflect the effective blockage geometry at the operational altitude slice.}

	\begin{figure}[!t]
		\centering
		\includegraphics[width=0.775\linewidth]{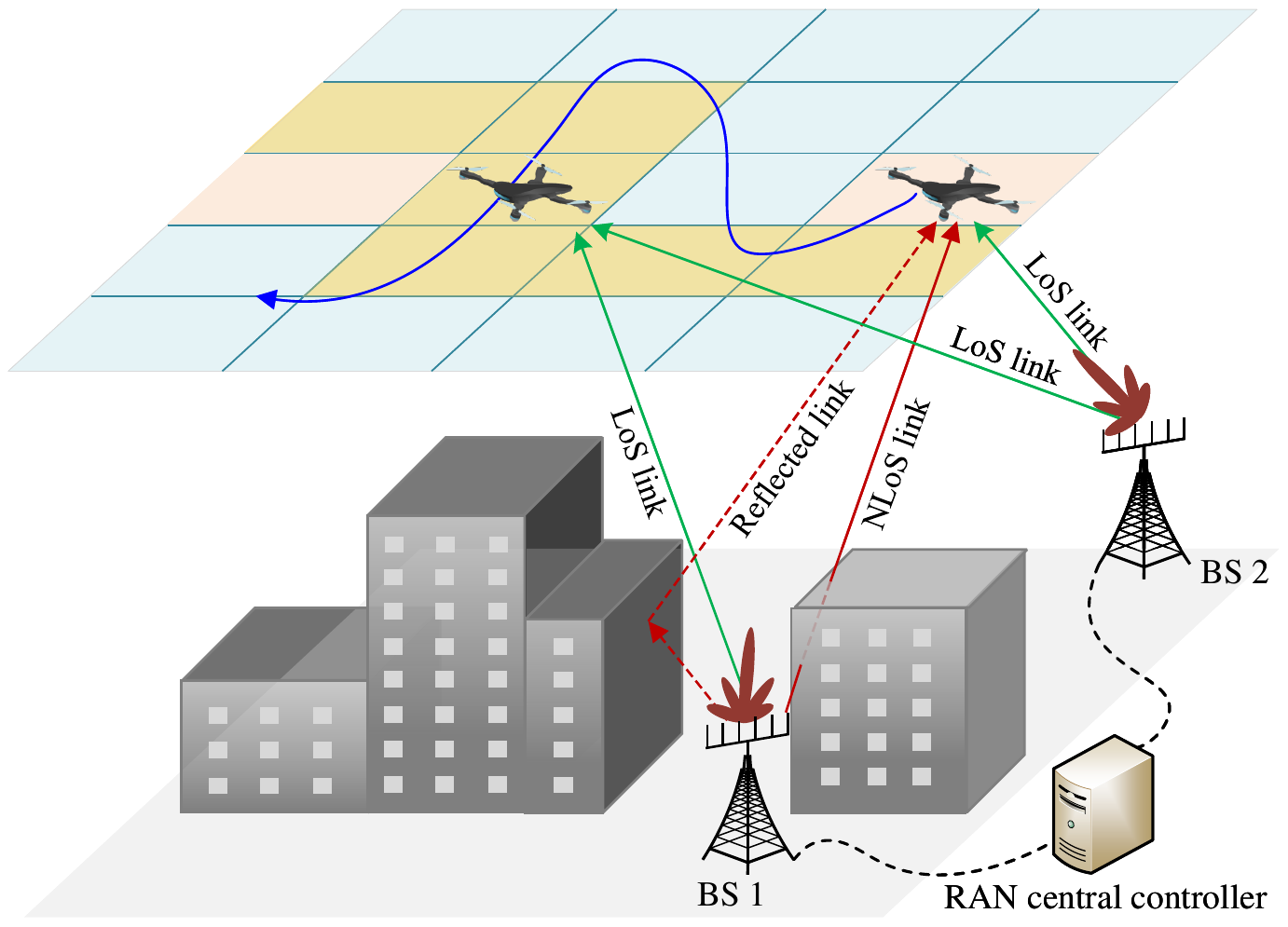}
		\caption{{Illustration of the edge-assisted UAV communication system in an urban scenario, where terrestrial BSs are interconnected with a {RAN central controller} via low-latency backhaul links.}
		}
		\label{fig:system_model}
		\vspace*{-0.4cm}
	\end{figure}

	The UAV conducts a mission along a continuous trajectory at a fixed altitude $z_{\rm h}$.{The resulting BeamCKM is a horizontal map at the operational altitude. A full three-dimensional CKM can be formed by stacking maps generated at multiple altitudes; extension to altitude-varying trajectories is left for future work.} The total flight period $T$ is discretized into $N$ equal time slots, defined as $\mathcal{N} \triangleq \{1,\dots,N\}$. The continuous horizontal location of the UAV at time slot $n$ is denoted by $\mathbf{q}[n] = [x[n], y[n]]^T \in{\mathcal{A}}$. The entire trajectory is defined as $\mathcal{Q} = \{\mathbf{q}[n]\}_{n=1}^N$, starting from $\mathbf{q}^{\rm start}$ and ending at $\mathbf{q}^{\rm end}$. To map the continuous UAV location $\mathbf{q}[n]$ to the discrete CKM grid, we define a spatial mapping function that outputs the corresponding grid indices: $\tilde{x}[n] = \lfloor x[n] / \Delta \kappa \rfloor$ and $\tilde{y}[n] = \lfloor y[n] / \Delta \kappa \rfloor$.

	\subsubsection{Communication model}
	Each BS $b \in \mathcal{B}$ is deployed at a fixed 3D location $\mathbf{q}_b = [x_b, y_b, z_b]^T$ and is equipped with a uniform linear array (ULA) consisting of $N_t$ antennas horizontally aligned parallel to the x-y plane\footnote{While this paper focuses on exploring the feasibility of CKM-based beamforming and handover using ULAs, the extension to uniform planar arrays (UPAs) is left for future work.}.
	The orientation angle of the array at BS $b$ is denoted by $\theta_{\rm ori}^{(b)}$. Let $\mathbf{w}_b \in \mathbb{C}^{N_t \times 1}$ denote the active continuous beamforming vector applied by BS $b$, satisfying the unit transmit power constraint $\|\mathbf{w}_b\|^2 = 1$. The received signal from BS $b$ at time slot $n$ is a superposition of $L$ multipath components, including LoS, reflections, and diffractions caused by the buildings, which can be expressed as
	\begin{align}\label{eq:channel_definition}
		Y_b(\mathbf{q}) = \sqrt{P} \sum_{l=1}^{L} \bar{h}_{b,l}(\mathbf{q}) \tilde{h}_{b,l}(\mathbf{q}) \mathbf{a}^H(\theta_l) \mathbf{w}_b X + Z,
	\end{align}
	where $P$ is the transmit power of each BS, $X$ is the transmitted data symbol with zero mean and normalized average power, and $Z \sim \mathcal{CN}(0, \sigma^2)$ is the additive white Gaussian noise (AWGN) with noise power $\sigma^2$. The vector $\mathbf{a}(\theta_l) \in \mathbb{C}^{N_t \times 1}$ denotes the array steering vector corresponding to the angle of departure (AoD) $\theta_l$ of the $l$-th path, capturing the physical spatial directivity of the antenna array.{The BSs operate on orthogonal time-frequency resources and therefore cause no inter-cell interference. The UAV is associated with one serving BS in each time slot; multi-BS interaction is limited to handover decisions and predictive beam preparation. Interference-aware simultaneous transmission is left for future work.}

	Modeled via ray-tracing \cite{yun2015access}, the BS-to-UAV downlink channel in \myrefeq{eq:channel_definition} exhibits large-scale fading $\bar{h}_{b,l}(\mathbf{q}) = \sqrt{\frac{G_{\rm t} G_{\rm r}}{{\rm PL}_{b,l}(\mathbf{q})}}$, where $G_{\rm t}$ and $G_{\rm r}$ denote the antenna gains of transmit and receive antenna, and the path loss is expressed as
	\begin{align}
		{\rm PL}_{b,l}(\mathbf{q}) = \prod_{i=1}^{I_R} |\Gamma_i|^{-2} \prod_{i=1}^{I_D} |\Lambda_i|^{-2} \left( \frac{4\pi d_{b,l}(\mathbf{q}) f}{c} \right)^2,
	\end{align}
	where $d_{b,l}(\mathbf{q})$ is the propagation distance, $f$ is carrier frequency, $c$ is light speed, $I_R$ is the number of reflections (each with material- and angle-dependent coefficient $\Gamma_i$, $0 < |\Gamma_i| < 1$), and $I_D$ is the number of diffractions (each with coefficient $\Lambda_i$, $0 < |\Lambda_i| < 1$). This formulation links ${\rm PL}_{b,l}$ to interaction counts and types, escalating with more reflections or diffractions.

	For small-scale fading, since ray-tracing determines all paths geometrically, $\tilde{h}_l(\mathbf{q})$ can be modeled deterministically for each distinct path as
	\begin{align}
		\tilde{h}_{b,l}(\mathbf{q}) = e^{j \phi_l(\mathbf{q})}, \,\, \phi_l(\mathbf{q}) = \frac{2\pi d_{b,l}(\mathbf{q})}{\lambda} +{2\pi f_{{\rm d},l} t} + \sum \phi_{i}^{(\cdot)},
	\end{align}
	where{$f_{{\rm d},l}$ denotes the Doppler shift of the $l$-th path induced by the UAV mobility, and} $\sum \phi_{i}^{(\cdot)}$ incorporates phase shifts{from reflections and diffractions}. In aerial communications, the rapid mobility and unavoidable sub-wavelength vibrations of UAVs induce severe Doppler shifts. This causes the instantaneous small-scale phase $\phi_l(\mathbf{q})$ to fluctuate drastically within milliseconds, which can be statistically modeled as uniformly distributed random variables over $[0, 2\pi]$ \cite{ tse2005fundamentals}. Consequently, we mathematically average out the microscopic transients and focus exclusively on the macroscopic expected channel characteristics, under the standard zero-mean unit-variance fast fading assumption $|\tilde{h}_{b,l}|^2 = 1$.

	The ground-truth expected channel power gain for a given location $\mathbf{q}$ and beamforming vector $\mathbf{w}_b$ is defined by averaging out the small-scale fading as
	\begin{align} \label{eq:expected_channel_gain}
		\mathcal{H}_b(\mathbf{q}, \mathbf{w}_b) = \mathbb{E}_{\tilde{h}} \left[ \left| \sum_{l=1}^{L} \bar{h}_{b,l}(\mathbf{q}) \tilde{h}_{b,l}(\mathbf{q}) \mathbf{a}^H(\theta_l) \mathbf{w}_b \right|^2 \right].
	\end{align}
	Note that $\mathcal{H}_b(\mathbf{q}, \mathbf{w}_b)$ inherently encapsulates the environmental blockages, pathloss, and the directional beamforming gain provided by $\mathbf{w}_b[n]$.
	Based on the formulation of expected channel gain in \myrefeq{eq:expected_channel_gain}, the downlink spectral efficiency for the UAV served by BS $b$ is given by
	\begin{align}\label{eq:approx_rate}
		R[n] = \log_2\left(1 + \frac{P \cdot {\mathcal{H}}_{b}(\mathbf{q}, \mathbf{w}_{b})}{\sigma^2}\right).
	\end{align}

	\begin{table}[!t]
		\caption{System notations}
		\label{tab:notation}
		\centering
		\renewcommand{\arraystretch}{1.1} % 稍微增大行距以便于阅读
		\setlength{\tabcolsep}{2mm}
		\begin{tabular}{l|l}
			\hline
			\textbf{Symbol} & \textbf{Description} \\
			\hline
			$\mathcal{B}$ & Set of terrestrial BSs \\
			$N$ & Number of time slots in the trajectory \\
			$N_t$ & Number of antennas in the uniform linear array (ULA) \\
			$P_{\max}$ & Maximum transmit power of the BS \\
			$\mathbf{w}_b[n]$ & Continuous beamforming vector of BS $b$ at slot $n$ \\
			$\mathbf{a}(\theta)$ & Antenna array steering vector at angle $\theta$ \\
			$L$ & Number of multipath components \\
			$\mathbf{q}[n]$ & Continuous position of the UAV at slot $n$ \\
			$\bar{h}_{b,l}(\cdot)$ & Large-scale channel fading of the $l$-th path from BS $b$ \\
			$\tilde{h}_{b,l}(\cdot)$ & Small-scale channel fading of the $l$-th path from BS $b$ \\
			$\sigma^2$ & Additive white Gaussian noise (AWGN) power \\
			$\mathcal{H}_b(\cdot, \cdot)$ & Ground-truth of expected channel gain mapping function \\
			$\hat{\mathcal{H}}_b(\cdot, \cdot)$ & Approximated channel gain within the Region of Interest \\
			$\mathbf{\Psi}_{\mathbf{w}_b[n]}^{(b)}$ & Ground-truth of BeamCKM for BS $b$ under beam $\mathbf{w}_b[n]$ \\
			$\hat{\mathbf{\Psi}}_{\mathbf{w}_b[n]}^{(b)}$ & Predicted BeamCKM generated by BeamCKMDiff \\
			$\mathcal{F}_\theta$ & Generative mapping surrogate function parameterized by $\theta$ \\
			{$\mathcal{A}$} & Continuous set of locations in the mission area \\
			{$\tilde{\mathcal{A}}$} & Discretized spatial grid of{$\mathcal{A}$} \\
			$\Delta \kappa$ & Horizontal spatial resolution of the CKM grid \\
			$\mathbf{E}$ & Physical environment topology matrix (building heights) \\
			$\mathbf{M}(\mathbf{q})$ & Spatial Region of Interest (RoI) mask for location $\mathbf{q}$ \\
			$\mathcal{Q}$ & Predefined continuous flight trajectory of the UAV \\
			$b[n]$ & Index of the associated serving BS at slot $n$ \\
			$r_{\rm e}$ & UAV position estimation error \\
			\hline
		\end{tabular}
		\vspace*{-0.25cm}
	\end{table}

	\subsubsection{Beam-Aware CKM and RoI Mapping}
	We define the ground-truth Beam-CKM as a 2D matrix $\mathbf{\Psi}_{\mathbf{w}_b[n]}^{(b)} \in \mathbb{R}^{N_1 \times N_2}$. The pixel value at grid $(i_x,j_y)$ corresponds to the expected channel gain $\mathcal{H}_b(\mathbf{q}_{i_x,j_y}, \mathbf{w}_b)$. Since acquiring the global ground-truth map via ray-tracing or exhaustive field measurements is prohibitive, our objective is to utilize a generative model to construct a predicted BeamCKM, denoted as
	\begin{align}
		\hat{\mathbf{\Psi}}_{\mathbf{w}_b}^{(b)} = \mathcal{F}_{\theta}(\mathbf{E}, \mathbf{q}_b, \mathbf{w}_b),
	\end{align}
	where $\mathcal{F}_{\theta}(\cdot)$ represents the proposed generative model, with $\theta$ denoting its trainable neural network parameters.

	Acquiring the precise coordinates of a highly mobile UAV is fundamentally challenging due to location errors.{The UAV coordinates are assumed to be obtained via onboard GNSS/inertial navigation feedback or network-side integrated sensing, where $r_{\rm e}$ denotes the maximum coordinate estimation error bound.} To guarantee robust connectivity, we define a Region of Interest (RoI) spatial mask $\mathbf{M}(\mathbf{q}) \in \{0, 1\}^{N_1 \times N_2}$, which is explicitly designed to encapsulate this spatial uncertainty bound. Mathematically, the $(i_x,j_y)$-th element of the mask is defined using an indicator function $\mathbb{I}(\cdot)$ as
	\begin{align} \label{eq:mask_def}
		[\mathbf{M}(\mathbf{q})]_{i_x,j_y} & = \\
		&  \mathbb{I}\left( \sqrt{(i_x \Delta \kappa - x[n])^2 + (j_y \Delta \kappa - y[n])^2} \leq r_{\rm e} \right). \notag
	\end{align}
	The predicted channel gain for location $\mathbf{q}$ is evaluated as the average value within the valid RoI, expressed as
	\begin{align} \label{eq:roi_gain}
		\hat{\mathcal{H}}_b(\mathbf{q}, \mathbf{w}_b) = \frac{\sum \left( \hat{\mathbf{\Psi}}_{\mathbf{w}_b}^{(b)} \odot \mathbf{M}(\mathbf{q}) \right)}{\sum \mathbf{M}(\mathbf{q})},
	\end{align}
	where $\odot$ denotes the Hadamard product. The key notations used in this paper are summarized in Table \ref{tab:notation}.

	\subsection{Problem Formulation}
	In this work, we aim to maintain a robust and high-capacity communication link by optimizing the continuous active beamforming vectors of BSs along the UAV trajectory. Note that while the time slot index $[n]$ was omitted in Section \ref{sec:system_model_a} for notational brevity, we explicitly reintroduce it here to capture the dynamic states across the $N$ flight slots. By directly utilizing the predicted spatial channel knowledge to evaluate the communication performance, the overall optimization problem to maximize the average spectral efficiency is formulated as
	\begin{align} \label{prob_final}
		{\rm (P1)}: &\max_{\{b[n], \mathbf{w}_{b}[n]\}_{n=1}^N} \quad \frac{1}{N} \sum_{n=1}^N \hat{R}[n] \\
		\text{s.t.} \,\,\, & \hat{R}[n] = \log_2\left(1 + \frac{P\cdot \hat{\mathcal{H}}_b(\mathbf{q}[n], \mathbf{w}_b[n])}{\sigma^2}\right), \, \forall n \in \mathcal{N} \tag{\ref{prob_final}a} \\
		& \|\mathbf{w}_{b}[n]\|^2 = 1, \quad \forall n \in \mathcal{N} \tag{\ref{prob_final}b} \\
		& b[n] \in \mathcal{B}, \quad \forall n \in \mathcal{N} \tag{\ref{prob_final}c}
	\end{align}
	where (\ref{prob_final}a) evaluates the achievable rate. Crucially, the channel gain $\hat{\mathcal{H}}_b$ in (\ref{prob_final}a) is spatially computed from the generated CKM $\hat{\mathbf{\Psi}}_{\mathbf{w}_b[n]}^{(b[n])} = \mathcal{F}_{\theta^*}(\cdot)$ via the RoI masking in \myrefeq{eq:roi_gain}, explicitly linking the beamforming variable $\mathbf{w}_b[n]$ to the objective without online measurements. Constraints (\ref{prob_final}b) and (\ref{prob_final}c) denote the normalized transmit-power budget and discrete BS selection, respectively.{Note that the rate is nondecreasing with respect to transmit power; hence an optimum of \myrefeq{prob_final} exhausts the power budget.}

	Problem \myrefeq{prob_final} is highly challenging due to the non-convex mapping from the complex-valued continuous beamforming manifold $\mathbf{w}$ to the generated spatial distribution $\hat{\mathbf{\Psi}}$. Conventional discrete pilot-sweeping methods fail to support continuous spatial optimization and incur unaffordable latency overhead. To effectively solve this problem, we decouple the original intractable formulation into two sequential sub-problems:
	\begin{enumerate}
		\item[-] \textit{BeamCKM construction}: The design and training of the generative framework, BeamCKMDiff, to construct a high-fidelity mapping function $\mathcal{F}_{\theta^*}(\cdot)$ that satisfies the spatial prediction requirement in (\ref{prob_final}f).
		\item[-] \textit{CKM-based beamforming and handover}: Leveraging the constructed BeamCKM as spatial prior, we optimize the beamforming vectors $\mathbf{w}_{b}[n]$ via gradient propagation and the BS association $b[n]$ to maximize the objective without online measurements.
	\end{enumerate}

	The detailed solutions for these two sub-problems are presented in Section III and Section IV, respectively.

	\section{Physics-Informed BeamCKMDiff}
	\label{sec:method}
\begin{figure*}[!t]
		\centering
		\includegraphics[width=0.92\linewidth]{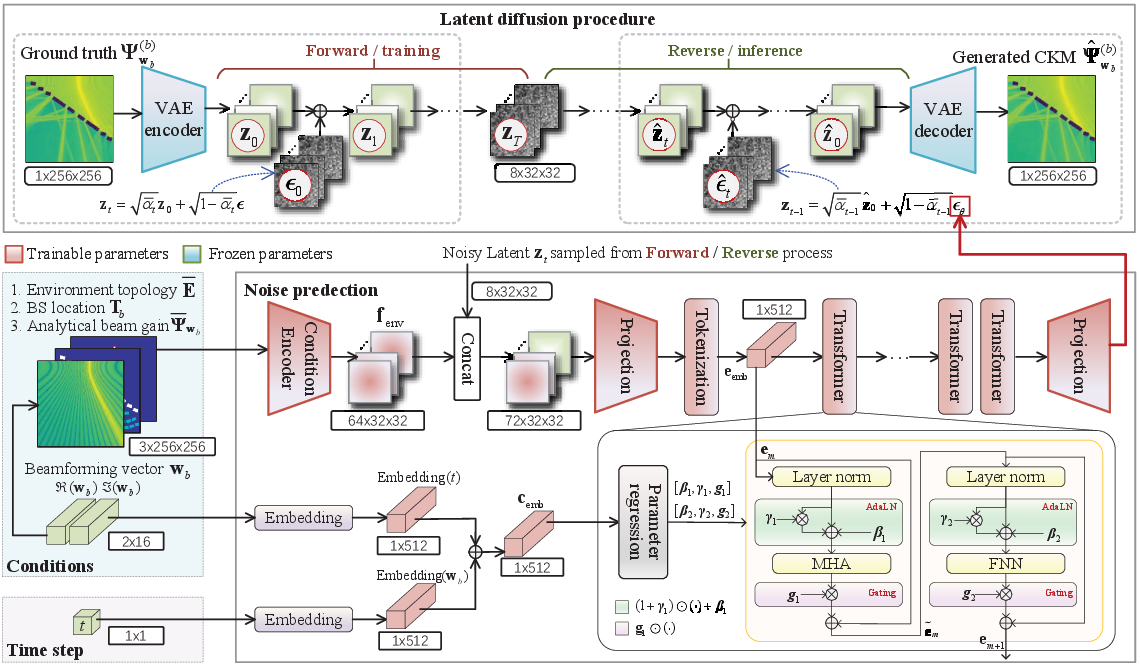}
		\caption{Architecture of BeamCKMDiff. Top: DeepVAE-based latent diffusion. Bottom: DiT noise predictor $\bm{\epsilon}_\theta$ driving the reverse inference (red arrow). Spatial conditions are concatenated with the noisy latent, while beam and time features are injected via adaLN-gating.}
		\label{fig:predicted_CKM_visualization}
		\vspace*{-0.2cm}
	\end{figure*}

	In this section, we present the design of the proposed BeamCKMDiff framework, which serves as the generative mapping surrogate $\mathcal{F}_\theta$. As illustrated in Fig. \ref{fig:predicted_CKM_visualization}, the framework consists of three key components: a deep variational autoencoder (DeepVAE) for perceptual compression, a deterministic condition encoder for environment topology and analytical beam prior extraction, and a DiT-based backbone featuring an adaLN mechanism for continuous beamforming vector embedding.

	\subsection{Perceptual Compression via DeepVAE}
	To avoid the prohibitive computational cost of pixel-space diffusion, we follow the latent approach in \cite{2021arXiv211210752R} and employ a DeepVAE to project the high-dimensional CKM into a compact latent space.
	The VAE comprises an encoder $\mathcal{E}$ and a decoder $\mathcal{D}$. Given a ground-truth global CKM $\mathbf{\Psi}_{\mathbf{w}_b}^{(b)} \in \mathbb{R}^{N_1 \times N_2}$, the encoder maps it to a latent Gaussian distribution $q(\mathbf{z}|\mathbf{\Psi}) = \mathcal{N}(\mathbf{z}; \bm{\mu}, \bm{\sigma}^2)$, yielding the sampled latent representation $\mathbf{z} \in \mathbb{R}^{h \times w \times D_{z}}$. The decoder $\mathcal{D}$ reconstructs the original map from $\mathbf{z}$. The training objective is to minimize the evidence lower bound (ELBO), combining a pixel-level reconstruction loss and a kullback-leibler (KL) divergence regularization term expressed as
	\begin{equation}
		\mathcal{L}_{\rm VAE} = \big\|\mathbf{\Psi} - \mathcal{D}(\mathbf{z})\big\|_2^2 + \lambda_{\rm KL} D_{\rm KL}\big(q(\mathbf{z}|\mathbf{\Psi}) \| \mathcal{N}(\mathbf{0}, \mathbf{I})\big).
	\end{equation}
	{Here, $\lambda_{\rm KL}>0$ weights the regularization term, and $D_{\rm KL}(\cdot\|\cdot)$ denotes the Kullback--Leibler divergence between the learned posterior $q(\mathbf{z}|\mathbf{\Psi})$ and the standard Gaussian prior $\mathcal{N}(\mathbf{0},\mathbf{I})$.} After training, the DeepVAE weights are frozen, and the subsequent generative process operates entirely within this efficient latent space.

	\subsection{Physics-Informed Condition Encoder}
	\label{subsec:condition_encoder}

	\subsubsection{Analytical Beam Prior}
	\label{subsec:analytical_prior}
	To guide the diffusion model with explicit electromagnetic propagation principles, we formulate an analytical beam prior that mathematically depicts the spatial energy distribution of the LoS path.
	For any specific grid index $(i_x, j_y)$ in the mission area, its corresponding 3D physical location at the UAV flight altitude is ${\mathbf{p}_{i_x,j_y}} = [i_x \Delta \kappa, j_y \Delta \kappa, z_h]^T$. Given the BS location $\mathbf{q}_b = [x_b, y_b, z_b]^T$ and its array orientation angle $\theta_{\rm ori}^{(b)}$, the global azimuth angle relative to the BS is calculated as
	\begin{align}
		\theta_{\rm global}^{(i_x,j_y)} = \arctan\left(\frac{i_x \Delta \kappa - x_b}{j_y \Delta \kappa - y_b}\right).
	\end{align}
	Consequently, the local azimuth AoD is
	\begin{align}
		\theta_{\rm local}^{(i_x,j_y)} = \theta_{\rm global}^{(i_x,j_y)} - \theta_{\rm ori}^{(b)}
		.	\end{align}
	Note that{the analytical formulation supplies only an LoS directivity prior for the array pattern. The generative model learns the NLoS multipath structure from data}.

	The elevation angle is computed as
	\begin{align}
		\theta_{\rm elev}^{(i_x,j_y)} = \arctan\left(\frac{z_h - z_b}{d_{\rm 2D}^{(i_x,j_y)}}\right),
	\end{align}
	where $d_{\rm 2D}^{(i_x,j_y)} = \sqrt{(i \Delta \kappa - x_b)^2 + (j \Delta \kappa - y_b)^2}$ denotes the 2D horizontal distance.
	Based on the 3D spatial geometry, the array factor $A(i_x,j_y, \mathbf{w}_b)$ for an $N_t$-antenna ULA applying the continuous beamforming vector $\mathbf{w}_b \in \mathbb{C}^{N_t \times 1}$ is explicitly derived as
	\begin{equation} \label{eq:array_factor}
		A(\mathbf{w}_b, i_x,j_y) = \sum_{n=0}^{N_t-1} [\mathbf{w}_b]^*_n e^{\left(-j 2\pi \frac{d}{\lambda} n \sin(\theta_{\rm local}^{(i_x,j_y)}) \cos(\theta_{\rm elev}^{(i_x,j_y)})\right)},
	\end{equation}
	where the superscript $(\cdot)^*$ denotes the complex conjugate
	operation, $d/\lambda$ denotes the normalized antenna spacing.
	Furthermore, the 3D propagation distance is $d_{\rm 3D}(i_x,j_y) = \sqrt{d_{\rm 2D}^{(i_x,j_y)\,2} + (z_{\rm h} - z_{\rm BS})^2}$. The free space path loss (FSPL) can be calculated as
	\begin{equation} \label{eq:fspl}
		{\rm FSPL}(i_x,j_y) = 20 \log_{10} \left( \frac{4\pi d_{\rm 3D}^{((i_x,j_y))} f_c}{c} \right),
	\end{equation}
	where $f_c$ is the carrier frequency and $c$ is the speed of light. Finally, the analytical beam gain prior map $\overline{\mathbf{\Psi}}_{\mathbf{w}_b} \in \mathbb{R}^{N_1 \times N_2}$ in dB scale can be constructed, where the pixel value at $(i_x,j_y)$ is given by
	\begin{equation} \label{eq:analytical_prior}
		\left[ \overline{\mathbf{\Psi}}_{\mathbf{w}_b} \right]_{i_x,j_y} = 10 \log_{10} \left( |A(\mathbf{w}_b, i_x,j_y)|^2 + \epsilon \right) - {\rm FSPL}(i_x,j_y),
	\end{equation}
	where $\epsilon$ is a negligibly small constant to ensure numerical stability.

	\subsubsection{Spatial Feature Extraction}
	Since the analytical prior map $\overline{\mathbf{\Psi}}_{\mathbf{w}_b}$ only captures LoS propagation, it must be integrated with the physical blockages. We normalize the raw building topology $\mathbf{E}$ by the UAV flight altitude $z_{\rm h}$ to reflect the relative obstruction severity. The normalized element at $(i_x,j_y)$ is expressed as
	\begin{equation}
		[\overline{\mathbf{E}}]_{i_x,j_y} = \min\left( \frac{[\mathbf{E}]_{i_x,j_y}}{z_{\rm h}}, 1 \right),
	\end{equation}
	where buildings exceeding $z_{\rm h}$ are capped at $1$, and lower buildings retain their proportional height.

	Additionally, we define a sparse binary transmitter map $\mathbf{T}_b$, where only the grid corresponding to the BS location $\mathbf{q}_b$ is set to $1$.
	These environmental maps are channel-wise concatenated with $\overline{\mathbf{\Psi}}_{\mathbf{w}_b}$ and fed into a ResNet-based condition encoder, denoted as $\mathcal{E}_{\rm env}$. Through strided convolutions and residual blocks, $\mathcal{E}_{\rm env}$ downsamples the input to match the latent space resolution, yielding the spatial condition tensor expressed as
	\begin{equation}
		\mathbf{f}_{\rm env} = \mathcal{E}_{\rm env}\big({\rm Concat}(\overline{\mathbf{E}}, \mathbf{T}_b, \overline{\mathbf{\Psi}}_{\mathbf{w}_b})\big)\in \mathbb{R}^{h \times w \times D_{\rm env}}.
	\end{equation}
	The condition encoder $\mathcal{E}_{\rm env}$ is optimized end-to-end with the DiT backbone.
	
%	le20253Dradiodiff
	\subsection{Continuous Beam Conditioning via DiT}
	The core generative engine of BeamCKMDiff is the DiT backbone, replacing the traditional CNN-based U-Nets used in prior works \cite{ZengYong_CKMDiff_arxiv, RadioDiff_TCCN}. We formulate the denoising process as a sequence-to-sequence modeling task, treating the noisy latent variable $\mathbf{z}_t$ at diffusion step $t$ and the spatial condition $\mathbf{c}_{\rm cond}$ as sequences of tokens.

	\subsubsection{Latent Feature Fusion}
	To integrate the spatial environmental constraints, the noisy latent $\mathbf{z}_t \in \mathbb{R}^{h \times w \times D_z}$ and the spatial condition tensor $\mathbf{f}_{\rm env} \in \mathbb{R}^{h \times w \times D_{\rm env}}$ are concatenated along the channel dimension. This fused representation is then partitioned into a sequence of patches and projected into a latent embedding via a patch embedding layer expressed as
	\begin{equation}
		\mathbf{e}_{\rm emb} = {\rm Proj}\left({\rm Concat}(\mathbf{z}_t, \mathbf{f}_{\rm env})\right) \in \mathbb{R}^{D_L \times D_{\rm h}},
	\end{equation}
	where $D_L$ is the sequence length and $D_{\rm h}$ denotes the hidden dimension of the transformer. We design $\mathbf{e}_{\rm emb}$ to serve as the primary sequence input to the first DiT block, carrying the joint information of the current diffusion state and the localized environmental topology.{The concatenated feature map is partitioned into non-overlapping $2\times2$ patches, producing $D_L$ tokens. The convolutional patch embedder $\mathrm{x\_emb}$ projects each patch to dimension $D_{\rm h}$, after which a learnable positional embedding preserves spatial order. Separate multilayer perceptrons map the diffusion step $t$ and the continuous beamforming vector $\mathbf{w}_b$ to condition embeddings. These embeddings are summed to obtain $\mathbf{c}_{\rm emb}$, which modulates every DiT block through adaLN.}

	\subsubsection{Beam-aware AdaLN}
	A core innovation of BeamCKMDiff is the dynamic modulation of the latent sequence using the continuous beamforming vector $\mathbf{w}_b$. This is achieved through a beam-aware adaLN mechanism, which treats the beam configuration and diffusion time-step as global control signals to steer the denoising process.

	First, the global joint embedding is synthesized by fusing the temporal and beam-specific descriptors expressed as
	\begin{equation}
		\mathbf{c}_{\rm emb} = {\rm Embedding}(t) + {\rm Embedding}(\mathbf{w}_b)\in \mathbb{R}^{D_{\rm h}},
	\end{equation}
	where $t$ is the diffusion time-step and $\mathbf{w}_b \in \mathbb{C}^{N_t\times 1}$ is the complex-valued beamforming vector. Note that both embeddings are implemented via linear projection layers to match the hidden dimension $D_{\rm h}$. For each DiT block, $\mathbf{c}_{\rm emb}$ is fed into a regression network to estimate six dimension-wise modulation parameters expressed as
	\begin{equation}
		[\bm{\beta}_1, \bm{\gamma}_1, \mathbf{g}_1, \bm{\beta}_2, \bm{\gamma}_2, \mathbf{g}_2] = \text{Linear}(\mathbf{c}_{\rm emb}),
	\end{equation}
	where all parameters $[\bm{\beta}, \bm{\gamma}, \mathbf{g}] \in \mathbb{R}^{D_{\rm h}}$ correspond to the scale, shift, and gating factors, respectively. Specifically, $\bm{\gamma}$ and $\bm{\beta}$ calibrate the feature distribution to reflect the spatial directivity of $\mathbf{w}_b$, while $\mathbf{g}$ scales the residual branch. The gating parameter $\mathbf{g}$ is initialized to zero, ensuring each block initially functions as an identity mapping for training stability.

	The interaction between the spatial tokens $\mathbf{e}_m \in \mathbb{R}^{D_L \times D_{\rm h}}$ and the joint conditioning $\mathbf{c}_{\rm emb}$ is executed as
	\begin{align}
		& \tilde{\mathbf{e}}_m = \mathbf{e}_m + \mathbf{g}_1 \odot {\rm MHA}\left( (1 + \bm{\gamma}_1) \odot {\rm LN}(\mathbf{e}_m) + \bm{\beta}_1 \right), \\
		& \mathbf{e}_{m+1} = \tilde{\mathbf{e}}_m + \mathbf{g}_2 \odot {\rm FFN}\left( (1 + \bm{\gamma}_2) \odot {\rm LN}(\tilde{\mathbf{e}}_m) + \bm{\beta}_2 \right),
	\end{align}
	where $m \in \{1,\dots,M_{\rm tran}\}$ is the layer of Transformer blocks, ${\rm MHA(\cdot)}$ is the multi-head attention (MHA) module, ${\rm FFN}(\cdot)$ is the feedforward neural network (FFN) module, $\mathbf{e}_0 = \mathbf{e}_{\rm emb}$, $\text{LN}(\cdot)$ denotes Layer normalization, and $\odot$ represents element-wise multiplication.
	%	By this design, $\mathbf{c}_{\rm emb}$ effectively steers the spatial features through the Transformer backbone, enabling the model to generate high-fidelity BeamCKM.

	\subsection{Diffusion Training and Inference}
	\label{subsec:training_inference}

	\subsubsection{Forward / Training Process}
	During training, the forward diffusion process progressively corrupts the encoded clean latent $\mathbf{z}_0 = \mathcal{E}(\mathbf{\Psi}_{\mathbf{w}_b}^{(b)})$ into a noisy state. At a randomly sampled time step $t \sim \mathcal{U}(1, T)$, the noisy latent $\mathbf{z}_t$ is constructed via
	\begin{equation}
		\mathbf{z}_t = \sqrt{\bar{\alpha}_t} \mathbf{z}_0 + \sqrt{1 - \bar{\alpha}_t} \bm{\epsilon}
	\end{equation}
	where $\bm{\epsilon} \sim \mathcal{N}(\mathbf{0}, \mathbf{I})$ is the ground-truth Gaussian noise injected at step $t$, and $\bar{\alpha}_t$ is the cumulative noise schedule parameter.

	To predict this injected noise at each step, the overall noise prediction mapping is denoted as $\bm{\epsilon}_\theta$. It integrates the previously detailed spatial features and the DiT backbone ${\rm DiT}_\theta(\cdot)$ into a unified function expressed as
	\begin{align} \label{eq:epsilon_theta}
		\bm{\epsilon}_\theta(\mathbf{z}_t, \mathbf{f}_{\rm env}, \mathbf{w}_b, t) & =  \, {\rm DiT}_\theta \Big( {\rm Proj}\big({\rm Concat}(\mathbf{z}_t, \mathbf{f}_{\rm env})\big), \\
		& {\rm Embedding}(t) + {\rm Embedding}(\mathbf{w}_b) \Big). \notag
	\end{align}
	The parameters $\theta$ are optimized by minimizing the mean squared error (MSE) between the true noise $\bm{\epsilon}$ and the predicted noise given by
	\begin{equation} \label{eq:dit_loss}
		\mathcal{L}_{\rm DiT} = \mathbb{E}_{t, \mathbf{z}_0, \bm{\epsilon}, \mathbf{w}_b} \Big[ \big\| \bm{\epsilon} - \bm{\epsilon}_\theta(\mathbf{z}_t, \mathbf{f}_{\rm env}, \mathbf{w}_{b}, t) \big\|_2^2 \Big].
	\end{equation}

	\subsubsection{Reverse Process via DDIM}
	To address the high latency of step-by-step markovian sampling in DDPMs, we adopt the DDIM \cite{song2022denoisingdiffusionimplicitmodels} scheme. At each reverse step $t$, the network $\bm{\epsilon}_\theta$ first estimates the clean latent $\hat{\mathbf{z}}_0$ based on the current noisy observation $\mathbf{z}_t$ expressed as
	\begin{equation} \label{eq:z0_est}
		\hat{\mathbf{z}}_0 = \frac{\mathbf{z}_t - \sqrt{1 - \bar{\alpha}_t} \bm{\epsilon}_\theta(\mathbf{z}_t, \mathbf{f}_{\rm env}, \mathbf{w}_b, t)}{\sqrt{\bar{\alpha}_t}}
	\end{equation}
	where $\bar{\alpha}_t$ is the cumulative noise schedule parameter. To control the stochasticity of the reverse process, DDIM generalizes the transition from $\mathbf{z}_t$ to $\mathbf{z}_{t-1}$ shown as
	\begin{equation} \label{eq:ddim_full}
		\mathbf{z}_{t-1} = \sqrt{\bar{\alpha}_{t-1}} \hat{\mathbf{z}}_0 + \sqrt{1 - \bar{\alpha}_{t-1} - \sigma_t^2} \cdot \bm{\epsilon}_\theta(\mathbf{z}_t, \mathbf{f}_{\rm env}, \mathbf{w}_b, t) + \sigma_t \bm{\epsilon}_t
	\end{equation}
	where $\bm{\epsilon}_t \sim \mathcal{N}(\mathbf{0}, \mathbf{I})$ is the standard Gaussian noise. The term $\sigma_t$ represents the standard deviation which is scaled by a stochasticity parameter $\eta \in [0, 1]$ given by
	\begin{equation} \label{eq:sigma_t}
		\sigma_t = \eta \sqrt{(1 - \bar{\alpha}_{t-1}) / (1 - \bar{\alpha}_t) \cdot (1 - \bar{\alpha}_t / \bar{\alpha}_{t-1})}.
	\end{equation}

	To maximize inference speed for real-time beamforming, we set $\eta = 0$ to eliminate the random noise term. In this case, $\sigma_t$ becomes zero, and the reverse process leads to a fully deterministic update expressed as
	\begin{equation} \label{eq:ddim_update}
		\mathbf{z}_{t-1} = \sqrt{\bar{\alpha}_{t-1}} \hat{\mathbf{z}}_0 + \sqrt{1 - \bar{\alpha}_{t-1}} \bm{\epsilon}_\theta(\mathbf{z}_t, \mathbf{f}_{\rm env}, \mathbf{w}_b, t).
	\end{equation}
	%	This deterministic mapping allows for significantly accelerated sampling using a non-Markovian sub-sequence of time steps while maintaining high construction accuracy.

	\section{CKM-based Beamforming and Handover}
	\label{sec:beamforming_handover}
	In this section, we address the continuous optimization of the beamforming vectors $\mathbf{w}_{b}$ and the BS association schedule $b[n]$ via the generatively constructed BeamCKM. The proposed framework translates the online spatial search into an analytical gradient-ascent procedure through the end-to-end architecture of BeamCKMDiff.

	\subsection{End-to-End Gradient Propagation}
	\label{subsec:end_to_end}

	The adopted DDIM framework with stochasticity parameter $\eta = 0$ constructs a deterministic transition path from the initial isotropic Gaussian noise $\mathbf{z}_T \sim \mathcal{N}(\mathbf{0}, \mathbf{I})$ to the clean latent $\mathbf{z}_0$.{Before online operation, $\mathbf{z}_T$ is drawn once using a fixed random seed. The same realization is retained throughout all beam-optimization iterations and candidate-BS evaluations. Consequently, the unrolled reverse process is a deterministic differentiable mapping with respect to $\mathbf{w}_b$, which removes sampling-induced gradient variation and enables variance-free link comparison.} At each reverse sampling step $t$, the transition function $\Phi_t$ is governed by the noise prediction network $\bm{\epsilon}_{\theta^*}$. The deterministic update is explicitly defined as
	\begin{align} \label{eq:ddim_deterministic}
		\mathbf{z}_{t-1} & = \Phi_t\left(\mathbf{z}_t, \mathbf{f}_{\rm env}(\mathbf{w}_b), \mathbf{w}_b\right) \\
		& = a_t \mathbf{z}_t + b_t \bm{\epsilon}_{\theta^*}\left(\mathbf{z}_t, \mathbf{f}_{\rm env}(\mathbf{w}_b), \mathbf{w}_b, t\right), \notag
	\end{align}
	where $a_t = \frac{\sqrt{\bar{\alpha}_{t-1}}}{\sqrt{\bar{\alpha}_t}}$ and $b_t = \sqrt{1 - \bar{\alpha}_{t-1}} - a_t \sqrt{1 - \bar{\alpha}_t}$ are constant schedule coefficients.

	Note that the continuous beamforming vector $\mathbf{w}_b$ steers the generation through a dual-path mechanism: it is explicitly injected into the DiT backbone ${\rm DiT}(\cdot)$ via the adaLN mechanism, and it inherently shapes the spatial condition tensor $\mathbf{f}_{\rm env}(\mathbf{w}_b)$ via the analytical beam prior. By unrolling the sequence of $T$ transition steps and decoding the final latent via the VAE decoder $\mathcal{D}(\cdot)$, the end-to-end generative mapping function $\mathcal{F}_{\theta^*}$ is formulated as a composition of differentiable mappings expressed as
	\begin{equation} \label{eq:surrogate_composition}
		\mathcal{F}_{\theta^*}(\mathbf{E}, \mathbf{q}_b, \mathbf{w}_b) = \mathcal{D} \circ \Phi_1 \circ \dots \circ \Phi_T \Big( \mathbf{z}_T; \, \mathbf{f}_{\rm env}(\mathbf{w}_b), \mathbf{w}_b \Big).
	\end{equation}

	Considering the spectral efficiency maximization in problem (P1) is mathematically equivalent to maximizing the expected channel gain. Let the target objective $\mathcal{J}(\mathbf{w}_b)$ evaluate this expected gain within the RoI mask $\mathbf{M}(\mathbf{q}[n])$. The objective function is directly derived from the output of the generative mapping, formulated as
	\begin{equation} \label{eq:obj_func}
		\mathcal{J}(\mathbf{w}_b) = \frac{1}{\|\mathbf{M}\|_1} \sum_{i=1}^{N_1} \sum_{j=1}^{N_2} \left( \left[ \mathcal{F}_{\theta^*}(\mathbf{E}, \mathbf{q}_b, \mathbf{w}_b) \right]_{i_x,j_y} \cdot [\mathbf{M}]_{i_x,j_y} \right).
	\end{equation}
	During the continuous beam optimization phase, the estimated physical location and its uncertainty bounds define the spatial mask $\mathbf{M}$. Crucially, since $\mathbf{M}$ is independent of the optimizable variable $\mathbf{w}_b$, it acts purely as a constant weight matrix. This mathematically decouples the boundary non-differentiability of the RoI mask from the beamforming optimization process.

	Thus, the CKM-based beamforming optimization sub-problem at time slot $n$ is explicitly formulated as
	\begin{align} \label{prob_sub1}
		\text{(P2)}: \max_{\mathbf{w}_b} \quad & \mathcal{J}(\mathbf{w}_b) \\
		\text{s.t.} \quad & \|\mathbf{w}_b\|^2 = 1. \tag{\ref{prob_sub1}a}
		%				& [\mathbf{M}(\mathbf{q}[n])]_{i_x,j_y} = \mathbb{I}\left( \sqrt{(i \Delta \kappa - x[n])^2 + (j \Delta \kappa - y[n])^2} \leq r_e \right), \, \forall i_x,j_y, n \tag{\ref{prob_sub1}b} \\
	\end{align}

	The total derivative of the objective $\mathcal{J}(\mathbf{w}_b)$ with respect to $\mathbf{w}_b$ is computed using the chain rule over the unrolled computational graph. The gradient flow from the objective back to the beamforming vector is derived as
	\begin{equation} \label{eq:chain_rule}
		\nabla_{\mathbf{w}_b} \mathcal{J}(\mathbf{w}_b) = \left( \frac{\mathrm{d} \mathbf{z}_0}{\mathrm{d} \mathbf{w}_b} \right)^T \nabla_{\mathbf{z}_0} \mathcal{D}(\mathbf{z}_0)^T \frac{\mathbf{M}}{\|\mathbf{M}\|_1}.
	\end{equation}
	The Jacobian matrix ${\mathrm{d} \mathbf{z}_0} / {\mathrm{d} \mathbf{w}_b}$ is accumulated by recursively applying the chain rule to the DDIM update sequence from $t=T$ down to $t=1$. Following the dual-path dependencies in \myrefeq{eq:ddim_deterministic}, the exact step-wise Jacobian update is given by
	\begin{equation} \label{eq:unrolled_grad}
		\frac{\mathrm{d} \mathbf{z}_{t-1}}{\mathrm{d} \mathbf{w}_b} = \underbrace{\frac{\partial \Phi_t}{\partial \mathbf{z}_t} \frac{\mathrm{d} \mathbf{z}_t}{\mathrm{d} \mathbf{w}_b}}_{\text{Historical gradient}} + \underbrace{\frac{\partial \Phi_t}{\partial \mathbf{w}_b}}_{\text{AdaLN gradient}} + \underbrace{\frac{\partial \Phi_t}{\partial \mathbf{f}_{\rm env}} \frac{\partial \mathbf{f}_{\rm env}}{\partial \mathbf{w}_b}}_{\text{Beam prior gradient}}.
	\end{equation}

	With the initial condition $\mathrm{d} \mathbf{z}_T / \mathrm{d} \mathbf{w}_b = \mathbf{0}$, the partial derivatives $\partial \Phi_t / \partial \mathbf{w}_b = b_t \partial \bm{\epsilon}_{\theta^*} / \partial \mathbf{w}_b$ and $\partial \Phi_t / \partial \mathbf{f}_{\rm env} = b_t \partial \bm{\epsilon}_{\theta^*} / \partial \mathbf{f}_{\rm env}$ explicitly capture the gradients propagated through the DiT backbone and the spatial condition encoder, respectively.
	Furthermore, the spatial prior gradient relies on the differentiability of the analytical beam prior $\overline{\mathbf{\Psi}}_{\mathbf{w}_b}$ constructed in \myrefeq{eq:analytical_prior}. By applying the chain rule, the spatial Jacobian is expanded as
	\begin{equation} \label{eq:spatial_jacobian}
		\frac{\partial \mathbf{f}_{\rm env}}{\partial \mathbf{w}_b} = \frac{\partial \mathcal{E}_{\rm env}}{\partial \overline{\mathbf{\Psi}}_{\mathbf{w}_b}} \frac{\partial \overline{\mathbf{\Psi}}_{\mathbf{w}_b}}{\partial \mathbf{w}_b}.
	\end{equation}
	The term ${\partial \mathcal{E}_{\rm env}}/{\partial \overline{\mathbf{\Psi}}_{\mathbf{w}_b}}$ is straightforwardly evaluated via the backpropagation of the CNN-based condition encoder. The term ${\partial \overline{\mathbf{\Psi}}_{\mathbf{w}_b}}/{\partial \mathbf{w}_b}$ bridges the physical electromagnetic propagation with the neural optimization. Since $\mathbf{w}_b$ is a complex-valued vector, we compute its gradient using Wirtinger calculus. Recall from \myrefeq{eq:analytical_prior} and denote the spatial steering vector at grid $(i_x,j_y)$ as $\mathbf{a}_{i_x,j_y}$; then the array factor is given by $A(\mathbf{w}_b, i_x,j_y) = \mathbf{a}_{i_x,j_y}^H \mathbf{w}_b$. The exact Wirtinger derivative of the prior map element with respect to the conjugate beamforming vector $\mathbf{w}_b^*$ is analytically derived as
	\begin{equation} \label{eq:wirtinger_prior}
		\nabla_{\mathbf{w}_b^*} \left[ \overline{\mathbf{\Psi}}_{\mathbf{w}_b} \right]_{i_x,j_y} = \frac{10}{\ln 10} \cdot \frac{\mathbf{a}_{i_x,j_y} \mathbf{a}_{i_x,j_y}^H \mathbf{w}_b}{|\mathbf{a}_{i_x,j_y}^H \mathbf{w}_b|^2 + \epsilon}.
	\end{equation}
	This closed-form derivative explicitly demonstrates that the physics-informed LoS prior imposes an analytical gradient field, dynamically guiding the diffusion model's spatial attention during the continuous beamforming optimization.

	The recursive formulation in \myrefeq{eq:unrolled_grad} mathematically guarantees that the generative BeamCKMDiff $\mathcal{F}_{\theta^*}$ acts as a continuous analytical function, facilitating end-to-end active beam optimization via reverse-mode automatic differentiation.

	\subsection{BeamCKMDiff-based Beamforming}
	We optimize the continuous beamforming vector $\mathbf{w}_b$ to maximize the spectral efficiency $R(\mathbf{w}_b)$ within the target RoI at time slot $n$. Because the achievable downlink rate $R(\mathbf{w}_b) = \log_2(1 + P \mathcal{J}(\mathbf{w}_b) / \sigma^2)$ is a monotonically increasing function of the channel gain $\mathcal{J}(\mathbf{w}_b)$, maximizing the spectral efficiency is mathematically equivalent to maximizing the spatial output of the generative mapping $\mathcal{F}_{\theta^*}$.

	To solve this non-convex optimization problem, we employ the successive convex approximation (SCA) method to iteratively update the beamforming vector. By denoting the iteration index as $l$, the gradient of the objective function with respect to the conjugate beamforming vector $\mathbf{w}_b^*$ is derived using the chain rule as
	\begin{equation}
		\nabla_{\mathbf{w}_b^*} R(\mathbf{w}_b) = \frac{P / \sigma^2}{\ln 2 \left(1 + P \mathcal{J}(\mathbf{w}_b) / \sigma^2\right)} \nabla_{\mathbf{w}_b^*} \mathcal{J}(\mathbf{w}_b).
	\end{equation}

	{Let $R^{(l)}[n]=R(\mathbf{w}_b^{(l)}[n])$ and $\mathbf{g}^{(l)}=\nabla_{\mathbf{w}_b^*}R(\mathbf{w}_b^{(l)}[n])$. Assuming that the gradient of $R$ is locally Lipschitz continuous, a step size $\mu_l$ satisfying the standard descent-lemma bound yields the concave quadratic minorant}
	{
	\begin{align} \label{eq:taylor_expansion}
		\check{R}^{(l)}& (\mathbf{w}_b) = R^{(l)}[n] \\
		& +\mathrm{Re}\!\left\{(\mathbf{g}^{(l)})^H(\mathbf{w}_b-\mathbf{w}_b^{(l)}[n])\right\}-\frac{1}{2\mu_l}\|\mathbf{w}_b-\mathbf{w}_b^{(l)}[n]\|_2^2. \notag
	\end{align}
	This surrogate is tight at $\mathbf{w}_b^{(l)}[n]$ and lower-bounds the local objective. {Maximizing it over the convex power ball $\|\mathbf{w}_b\|_2\leq1$ is equivalent to a projected-gradient update}. The intermediate beamforming vector is therefore updated as
	\begin{equation}
		\tilde{\mathbf{w}}_b^{(l+1)}[n] = \mathbf{w}_b^{(l)}[n] + \mu_l \mathbf{g}^{(l)},
	\end{equation}
	Here, $\mu_l=\mu_0\eta^{l}$ is the predefined step size. The scalar $\tilde{w}^{(l+1)}_{b,0}$ is the first element of $\tilde{\mathbf{w}}_b^{(l+1)}[n]$. Projection onto the power ball and removal of the global phase ambiguity give
	\begin{equation}\label{eq:inter_normalize}
		\mathbf{w}_b^{(l+1)}[n] = \frac{\tilde{\mathbf{w}}_b^{(l+1)}[n]\exp\!\left(-j\angle\tilde{w}^{(l+1)}_{b,0}\right)}{\max\{1,\|\tilde{\mathbf{w}}_b^{(l+1)}[n]\|_2\}}.
	\end{equation}
	
	Under the standard SCA conditions of surrogate tightness, gradient consistency, and appropriate step-size selection, the objective sequence is nondecreasing and every limit point is stationary \cite{Scutari2017SCA}. Equation \myrefeq{eq:inter_normalize} is thus the exact solution of the proximal SCA subproblem. Hierarchical clustering can extend the optimization to large networks by decomposing candidate BSs or UAVs into local groups \cite{cluster2025}.}

	\subsection{BeamCKMDiff-Based Dual-BS Handover}
	\label{subsec:handover}

\begin{figure}[!t]
		\centering
		\includegraphics[width=0.99\linewidth]{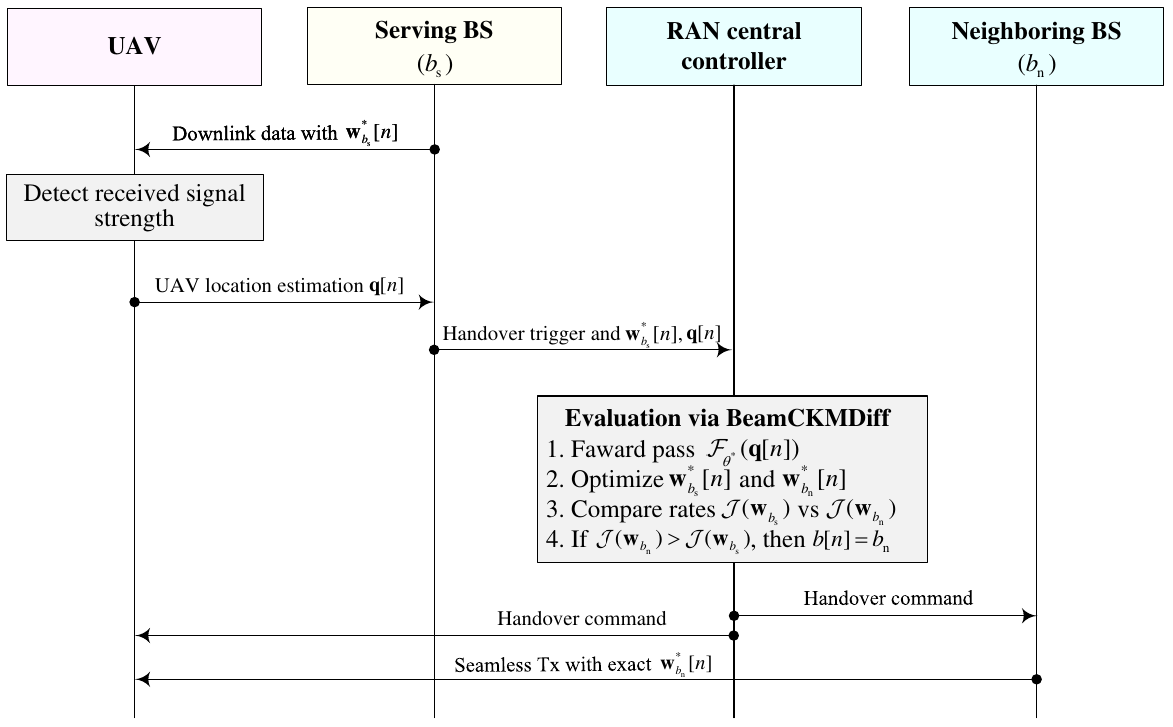}
		\caption{{Signaling procedure of the proactive dual-BS handover with RAN central controller evaluation.}
		}
		\label{fig:msc_handover}
		\vspace*{-0.2cm}
	\end{figure}

	Due to the high mobility of UAVs, the spatial channel conditions change rapidly, causing the communication link provided by a single serving BS to easily become outdated. To maintain robust continuous connectivity and overcome the coverage limitations of a single BS, we propose a BeamCKMDiff-based dual-BS handover scheme. As illustrated in Fig. \ref{fig:msc_handover}, this framework leverages the generative mapping $\mathcal{F}_{\theta^*}(\cdot)$ to proactively evaluate the potential link quality of candidate BSs, replacing exhaustive physical pilot sweeping with a model-based analytical evaluation.

	At any given time slot $n$, the UAV is served by the current BS, denoted as $b_{\rm s}$, using the optimized continuous beamforming vector $\mathbf{w}_{b, {\rm s}}^*[n]$. The UAV continuously detects the received signal strength ${\rm RSS}_{b, {\rm s}}[n]$. If the measured signal drops below a predefined handover trigger threshold $\gamma_{\rm th}$ or drops over $\gamma_{\rm drop}$, the system initiates the handover evaluation procedure.
	Instead of forcing the neighboring BS, denoted as $b_{\rm n}$, to transmit physical scanning pilots, the UAV simply reports its real-time location estimation $\mathbf{q}[n]$ to the serving BS. Subsequently, the proposed framework invokes the BeamCKMDiff model{at the {RAN central controller}} to perform a virtual evaluation. Specifically, the forward pass $\mathcal{F}_{\theta^*}(\mathbf{q}[n])$ is executed to generate the spatial prior. Based on this, we apply the projected gradient ascent formulated in Section IV-B to optimize the continuous beamforming vectors $\mathbf{w}_{b, {\rm s}}^*[n]$ and $\mathbf{w}_{b, {\rm n}}^*[n]$ simultaneously.

	The expected received signal strength for both BSs are computed accordingly as $\mathcal{J}(\mathbf{w})_{b, {\rm s}}$ and $\mathcal{J}(\mathbf{w})_{b, {\rm n}}$. The dual-BS association decision is determined by comparing their respective achievable rates expressed as
	\begin{equation} \label{eq:handover_decision}
		b[n] =
		\begin{cases}
			b_{\rm n}, & \text{if } \mathcal{J}(\mathbf{w}_{b, {\rm n}}) > \mathcal{J}(\mathbf{w}_{b, {\rm s}}) + \Delta_{\rm hyst}, \\
			b_{\rm s}, & \text{otherwise},
		\end{cases}
	\end{equation}
	where $\Delta_{\rm hyst}$ is a hysteresis margin introduced to prevent the ping-pong effect caused by transient environmental noise.
	Once the condition $\text{if } \mathcal{J}(\mathbf{w}_{b, {\rm n}}) > \mathcal{J}(\mathbf{w}_{b, {\rm s}}) + \Delta_{\rm hyst}$ is satisfied, the handover decision $b[n] = b_{\rm n}$ is made. The serving BS then dispatches the handover command to both the UAV and the neighboring BS. By utilizing the model-generated spatial prior, the target BS $b_{\rm n}$ acquires the exact optimal continuous beamforming vector proactively.

	{As shown in Fig.~\ref{fig:msc_handover}, the RAN central controller coordinates handover decisions. Upon a signal-degradation trigger, it evaluates candidate links and computes the target beam $\mathbf{w}_{b_{\rm n}}^*[n]$ in parallel. Once the handover condition is met, the target BS activates the pre-configured beam without physical pilot sweeping. Additional BSs enlarge only the candidate set and do not change the procedure.}
	
	\begin{algorithm}[!t]
		\caption{Proposed CKM-based Beamforming and Handover}
		\label{alg:overall_algo}
		\SetAlgoLined
		\textbf{Input}: Trajectory $\mathcal{Q}$, topology $\mathbf{E}$, BS set $\{\mathbf{q}_b\}$, bounds $r_{\rm e}$, step $\mu$, thresholds $\gamma_{\rm th}$, $\gamma_{\rm drop}$, $\Delta_{\rm hyst}$. \\
		Pre-train BeamCKMDiff $\mathcal{F}_{\theta^*}(\cdot)$ via \eqref{eq:dit_loss}. \\
		Initialize $\mathbf{w}_{b, {\rm s}}^*[0]$ with an arbitrary feasible unit-power vector. \\
		\For{$n = 1, \dots, N$}{
			\tcp{Step 1: State Monitoring}
			Obtain location $\mathbf{q}[n]$, construct RoI mask $\mathbf{M}(\mathbf{q}[n])$ via \eqref{eq:mask_def}. \\
			Measure downlink received signal strength ${\rm RSS}_{b, {\rm s}}[n]$ from the serving BS $b_s$. \\
			
			\tcp{Step 2: CKM-based Beamforming}
			Initialize $\mathbf{w}_{b, {\rm s}}^{(0)}[n] = \mathbf{w}_{b, {\rm s}}^*[n-1]$ for cold-start acceleration. \\
			\For{$l = 0, \dots, L_{\max}-1$}{
				Evaluate expected channel gain $\mathcal{J}(\mathbf{w}_{b, {\rm s}}^{(l)})$ via \eqref{eq:obj_func}. \\
				Compute objective gradient $\nabla_{\mathbf{w}_{b, {\rm s}}^*} R$ using chain rule \eqref{eq:chain_rule}, integrating spatial prior \eqref{eq:wirtinger_prior} and DiT Jacobian \eqref{eq:unrolled_grad}. \\
				Update and phase-aligned normalize: \\
				$\mathbf{w}_{b, {\rm s}}^{(l+1)} = \frac{\tilde{\mathbf{w}}_{b, {\rm s}}^{(l+1)} \exp(-j \angle \tilde{w}_{b_{\rm s}, 0}^{(l+1)})}{\|\tilde{\mathbf{w}}_{b_s}^{(l+1)}\|_2}$, where $\tilde{\mathbf{w}}_{b, {\rm s}}^{(l+1)} = \mathbf{w}_{b, {\rm s}}^{(l)} + \mu \nabla_{\mathbf{w}_{b, {\rm s}}^*} R$.
			}
			Set optimal serving beam: $\mathbf{w}_{b, {\rm s}}^*[n] \leftarrow \mathbf{w}_{b, {\rm s}}^{(L_{\max})}[n]$. \\
			
			\tcp{Step 3: CKM-based Handover}
			\If{${\rm RSS}_{b, {\rm s}}[n] < \gamma_{\rm th}$ \textbf{or} ${\rm RSS}_{b, {\rm s}}[n-1] - {\rm RSS}_{b, {\rm s}}[n] > \gamma_{\rm drop}$}{
				Obtain $\mathbf{w}_{b, {\rm n}}^*[n]$ via \textbf{Step 2}. \\
				Calculate and compare expected rates $R_{b, {\rm n}}$ and $R_{b, {\rm s}}$. \\
				\If{$R_{b, {\rm n}} > R_{b, {\rm s}} + \Delta_{\rm hyst}$}{
					Execute handover: $b_{\rm s} \leftarrow b_{\rm n}$.
				}
			}
		}
		\textbf{Output}: Optimized $\{\mathbf{w}_{b[n]}^*[n]\}$ and $\{b[n]\}$.
	\end{algorithm}
	
	\subsection{Overall Algorithm}
	The overall BeamCKMDiff framework is summarized in Algorithm \ref{alg:overall_algo}. After offline pre-training of $\mathcal{F}_{\theta^*}(\cdot)$, it executes an end-to-end continuous beamforming procedure. In each slot, the serving beam $\mathbf{w}_{b, {\rm s}}[n]$ is optimized via unrolled gradient ascent over the location-conditioned spatial prior. To prevent outages, severe signal degradation triggers a virtual evaluation for candidate BS $b_{\rm n}$. Leveraging the generated prior, the target BS proactively acquires its optimal beam without physical pilot scanning, achieving seamless handover.

	{In deployment, an RAN central controller connected to the BSs performs BeamCKMDiff inference and SCA optimization. In each time slot, the UAV reports its estimated horizontal coordinates and one RSS value; it performs no neural-network computation. Centralized candidate-link evaluation eliminates inter-BS CSI exchange. The model is trained once offline and accepts the target topology matrix at inference, so a new environment does not require online retraining.}

	{The online cost comprises analytical-prior generation, condition encoding, reverse-diffusion forward and backward propagation, and VAE decoding. One SCA iteration requires $\mathcal{O}(N_1N_2N_t)$ operations for the LoS prior and $\mathcal{O}(L_{\rm CNN}N_1N_2K^2)$ for condition encoding. Let $C_{\rm DiT}=\mathcal{O}(M_{\rm tran}D_L^2D_{\rm h})$ denote the cost of one DiT evaluation at one DDIM step. Each forward or backward traversal then costs $\mathcal{O}(T_{\rm step}C_{\rm DiT})$. The VAE decoder costs $\mathcal{O}(L_{\rm VAE}hwf^2)$ and is invoked once after convergence. For a mission with $N$ time slots and $N_{\rm HO}$ handover triggers, each involving $|\mathcal{B}_{\rm cand}|$ candidate BSs, the dominant mission-wide cost is $\mathcal{O}((N+N_{\rm HO}|\mathcal{B}_{\rm cand}|)L_{\max}T_{\rm step}C_{\rm DiT})$.} {Table \ref{tab:complexity} summarizes all terms. By executing this optimization at the grid-powered RAN central controller, the computational cost is paid on the ground to eliminate physical beam scanning over the air interface.}

	\begin{table*}[!t]
		\caption{Complexity Breakdown of Algorithm Components}
		\label{tab:complexity}
		\centering
		\setlength{\tabcolsep}{5mm}
		\renewcommand{\arraystretch}{1.15}
		\begin{tabular}{l | l}
			\Xhline{1.1pt}
			\textbf{Component} & \textbf{Complexity Expression} \\
			\hline
			Analytical Prior Generation & $\mathcal{O}(N_1 N_2 N_t)$ per step \\
			Condition Encoder & $\mathcal{O}(L_{\rm CNN} \cdot N_1 N_2 \cdot K^2)$ per step \\
			{DiT Forward Pass} &{$\mathcal{O}(T_{\rm step} \cdot M_{\rm tran} \cdot D_L^2 D_{\rm h}) \triangleq \mathcal{O}(T_{\rm step} C_{\rm DiT})$} \\
			{Unrolled Graph Backward Pass} &{$\mathcal{O}(T_{\rm step} C_{\rm DiT})$} \\
			VAE Decoder & $\mathcal{O}(L_{\rm VAE} \cdot h w \cdot f^2)${(invoked once upon convergence)} \\
			\hline
			\textbf{{Online SCA per BS per Slot}} &{$\mathcal{O}\!\left(L_{\max}\left(T_{\rm step}C_{\rm DiT}+L_{\rm CNN}N_1N_2K^2+N_1N_2N_t\right)+L_{\rm VAE}hwf^2\right)$} \\
			{\textbf{Total Mission Complexity}} &{$\mathcal{O}\left( (N + N_{\rm HO} |\mathcal{B}_{\rm cand}|) \cdot L_{\max} T_{\rm step} C_{\rm DiT} \right)$} \\
			\Xhline{1.1pt}
		\end{tabular}
		\vspace{-0.15cm}
	\end{table*}

	\section{Simulation Results}
	\label{sec:simulation}

	%	This section evaluates the performance of the proposed BeamCKMDiff framework and the BeamCKMDiff-based beamforming and handover algorithms. We begin by introducing the simulation setup, including the dataset construction and parameter configurations. Subsequently, we compare our proposed methods against baseline schemes to assess their accuracy in beam-aware CKM construction and their communication performance.

	\subsection{Simulation Settings}

	The BeamCKM dataset is constructed from 110 geo-referenced urban topologies extracted from OpenStreetMap \cite{osm_planetdump_2017}, divided into 100 environments for training and 10 for testing. For unseen testing environments, only the 3D environmental topology and transceiver coordinates are known a priori; channel gains are entirely inferred by the model. Each scenario covers a $512 \text{ m} \times 512 \text{ m}$ area, discretized into a grid with a spatial resolution of $\Delta \kappa = 2 \text{ m}$. BSs equipped with a $32 \times 1$ ULA are deployed at 10 random locations at a height of $10 \text{ m}$. To generate diverse spatial propagation characteristics, 10 fully digital random beamforming vectors are applied per deployment. Although these random phases lack practical directivity, they are deliberately employed to prevent the generative model from overfitting to restricted discrete codebooks, thereby ensuring a smooth and generalized mapping that provides accurate gradient flows for the continuous SCA optimization. The continuous beamforming vectors are power-normalized with uniformly distributed relative phases, mathematically expressed as
	\begin{align}
		\mathbf{w} = \frac{1}{\sqrt{N_t}} [e^{j\phi_1}, e^{j\phi_2}, \dots, e^{j\phi_{N_t}}]^T,
	\end{align}
	where $\phi_n \sim \mathcal{U}(0, 2\pi)$. This configuration yields a comprehensive dataset comprising $10,000$ training samples and $1,000$ testing samples. High-fidelity channel gains are generated using the NVIDIA Sionna ray-tracing engine \cite{hoydis2022sionna}, operating at a carrier frequency of $28 \text{ GHz}$. The Sionna module is configured with $10^9$ maximum rays, up to 3 reflections, and diffraction enabled.{For each BS deployment, Sionna generates the ground-truth BeamCKMs of the ten sampled continuous beams in one offline ray-tracing run. Exhaustive sampling is infeasible because the beam space is continuous. Random phase vectors provide broad training coverage, while independently sampled test beams assess interpolation to unseen vectors. By contrast, measurement-based BeamCKM construction requires a separate acquisition for each candidate beam \cite{wang2025beamckm}. After training, BeamCKMDiff generates a map for any continuous beam without further measurements or ray tracing, amortizing the offline cost over subsequent queries.}

	\begin{table}[!t]
		\renewcommand{\arraystretch}{1.15}
		\centering
		\setlength{\tabcolsep}{3mm}
		\caption{Model Architecture Configurations}
		\label{table_model_summary}
		\begin{tabular}{l|p{5.8cm}}
			\Xhline{1.1pt}
			\textbf{Module} & \textbf{Core Components \& Hyperparameters} \\ \hline
			\textbf{VAE} & \textit{In}: $1 \times 256^2$, \textit{Latent}: $D_z \times h \times w$ ($8 \times 32^2$) \\
			\quad - \textit{Encoder} & Input $\xrightarrow{\text{Conv}}$ [Downsample $\to$ \textbf{ResNet}]$_{\times 3}$ \\
			& Channels: $1 \to [64 \to 128 \to 256] \to 8$ \\
			& \texttt{map\_latent}: Conv2d(256, 8, $3\times3$) \\
			\quad - \textit{Decoder} & \textit{Out}: $1 \times 256^2$ \newline
			$\mathbf{z} \xrightarrow{\text{Conv}}$ [ResNet $\to$ Upsample $\to$ Conv]$_{\times 3}$ \\
			& Channels: $8 \to 256 \to [128 \to 64 \to 32] \to 1$ \\
			& \texttt{final}: Conv2d(32, 1, $3\times3$) $\to$ Sigmoid \\ \hline
			\textbf{ResNetBlock} & \texttt{gn1}: GroupNorm(8, $C_{in}$) $\to$ SiLU \\
			& \texttt{conv1}: Conv2d($C_{in}, C_{out}, 3\times3$) \\
			& \texttt{gn2}: GroupNorm(8, $C_{out}$) $\to$ SiLU \\
			& \texttt{conv2}: Conv2d($C_{out}, C_{out}, 3\times3$) \\ \hline
			\textbf{Cond. Enc.} & \textit{In}: $3 \times 256^2$, \textit{Out}: $D_{\text{env}} \times h \times w$ ($64 \times 32^2$) \\
			& \texttt{head}: Conv2d(3, 32, $3\times3$) \\
			& Body: [ResNetBlock $\to$ Downsample]$_{\times 3}$ \\
			& Channels: $32 \to [32 \to 64 \to 128]$ \\
			& \texttt{out}: Zero-init Conv2d(128, 64, $3\times3$) \\ \hline
			\textbf{BeamDiT} & \textit{In}: $(D_z + D_{\text{env}}) \times h \times w$ ($72 \times 32^2$) \\
			\quad - \textit{Embedders} & \texttt{x\_emb}: Conv2d(72, $D_{\text{h}}=512$, $2\times2$, stride=2) \\
			& \texttt{pos\_emb}: Learnable ($1 \times D_L \times D_{\text{h}}$) \newline ($D_L=256, D_{\text{h}}=512$) \\
			& \texttt{t/w\_emb}: MLP(Linear $\to$ SiLU $\to$ Linear) \\
			\quad - \textit{Backbone} & \textbf{Depth}: 12, \textbf{Heads}: 8, \textbf{Hidden} ($D_{\text{h}}$): 512 \\
			\quad - \textit{DiT Block} & \texttt{adaLN}: Regresses $[\bm{\gamma}_1, \bm{\beta}_1, \bm{g}_1, \bm{\gamma}_2, \bm{\beta}_2, \bm{g}_2]$ \\
			& \texttt{attn}: MultiheadAttn (dim=512, heads=8) \\
			& \texttt{mlp}: Linear(512 $\to$ 2048) $\to$ GELU $\to$ Linear(512) \\
			\quad - \textit{Final} & \texttt{norm}: LayerNorm + adaLN (Scale $\gamma$, Shift $\beta$) \\
			& \texttt{linear}: Linear($D_{\text{h}}$, $2^2 \times D_z$) $\to$ Linear(512, 32) \\
			\Xhline{1.1pt}
		\end{tabular}
		\vspace*{-0.2cm}
	\end{table}

	\subsubsection{Evaluation and System Configurations}
	During the forward diffusion process, noise is injected into the latent representation over $T=500$ timesteps, governed by a linear variance schedule ranging from $\beta_1 = 1 \times 10^{-4}$ to $\beta_T = 0.02$. The model parameters are optimized using the AdamW optimizer with a batch size of $4$. The VAE and BeamDiT modules are trained over $5000$ and $1,000$ epochs with learning rates of $10^{-3}$ and $10^{-4}$, respectively. The specific hyperparameter configurations are detailed in Table \ref{table_model_summary}. For the inference phase, the DDIM sampler is employed to accelerate generation, utilizing $T_{\text{step}}=5$ for optimal visual quality and fast beamforming.{All architectural and optimization hyperparameters are fixed a priori from standard generative-model configurations. No test environment is used for model selection or parameter tuning; evaluation is strictly zero-shot and out of distribution.}

	{The downlink BSs serve a UAV flying at $z_{\rm h}=100$ m and transmit at $P=40$ dBm. The $100$ MHz bandwidth conforms to the FR2 channel bandwidth specified in 3GPP TS 38.104 \cite{3gpp_ts_38_104}. A thermal-noise density of $-174$ dBm/Hz and a $7$ dB receiver noise figure yield an effective noise power of approximately $-87$ dBm. We set the outage threshold to $-80$ dBm, which provides a $7$ dB margin above the noise floor. Handover evaluation is triggered when the serving-link RSS falls below $\gamma_{\rm th}=-70$ dBm or drops by $\gamma_{\rm drop}=3$ dB. A $3$ dB hysteresis margin suppresses ping-pong handovers.}
	The SCA optimization parameters are configured with a maximum number of iterations $L_{\max} = 30$, an initial step size $\mu_0 = 0.1$, and an exponential decay rate $\eta = 0.85$ to ensure rapid convergence. All procedures are executed on an NVIDIA GeForce RTX 4090 GPU.

	\begin{figure*}[!t]
		\centering
		\subfloat{\includegraphics[width=1.35in]{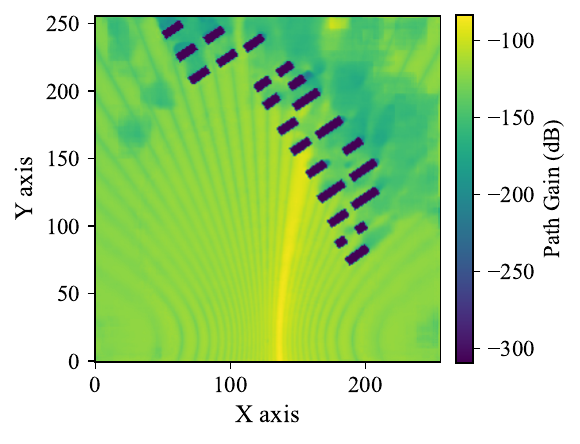}}
		\subfloat{\includegraphics[width=1.35in]{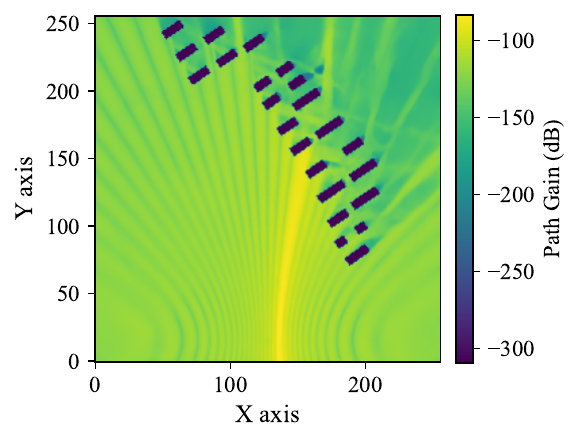}}
		\subfloat{\includegraphics[width=1.35in]{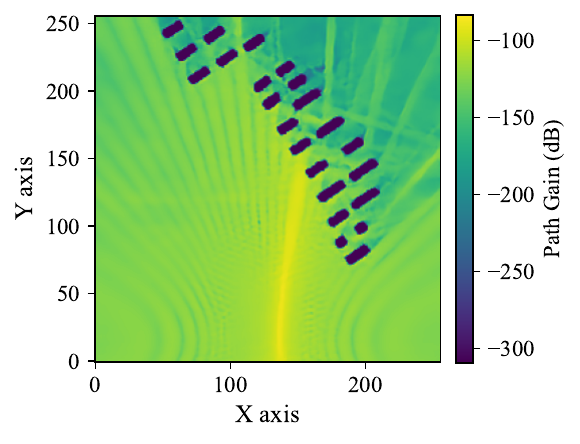}}
		\subfloat{\includegraphics[width=1.35in]{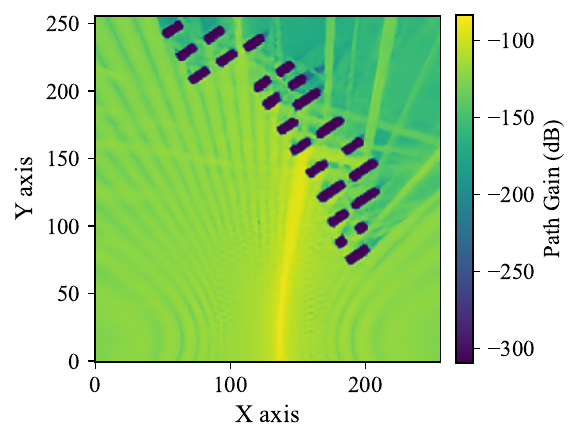}}
		\subfloat{\includegraphics[width=1.35in]{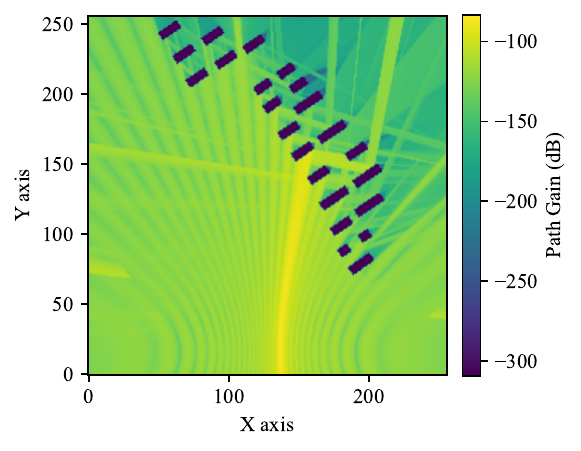}}
		\\
		\subfloat{\includegraphics[width=1.35in]{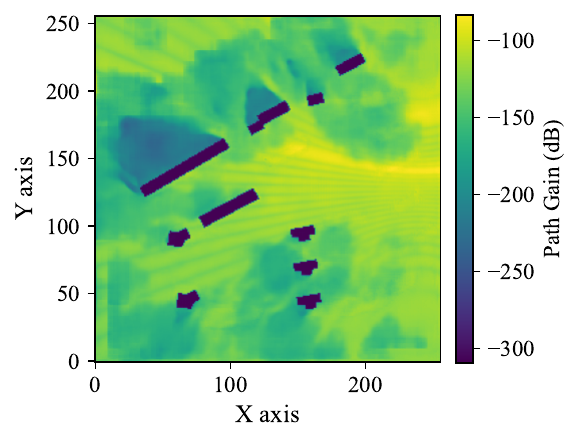}}
		\subfloat{\includegraphics[width=1.35in]{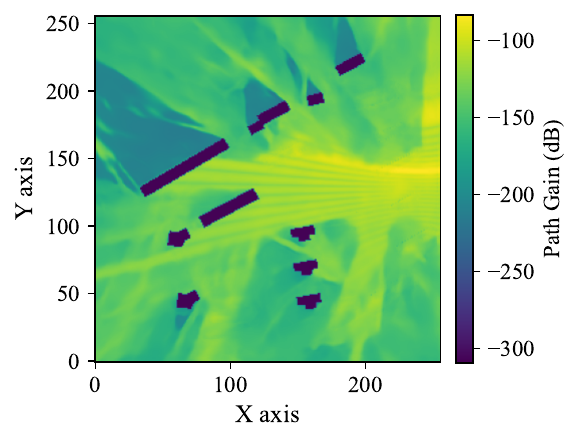}}
		\subfloat{\includegraphics[width=1.35in]{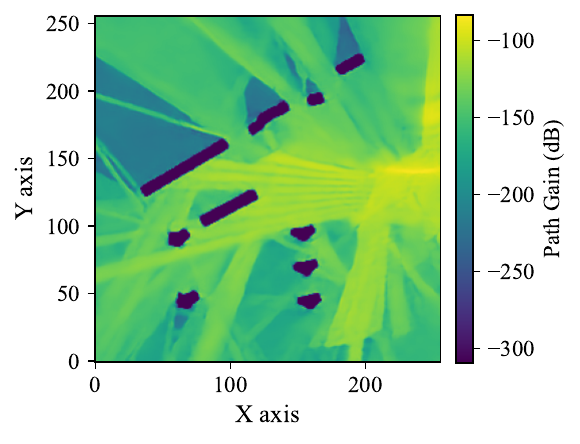}}
		\subfloat{\includegraphics[width=1.35in]{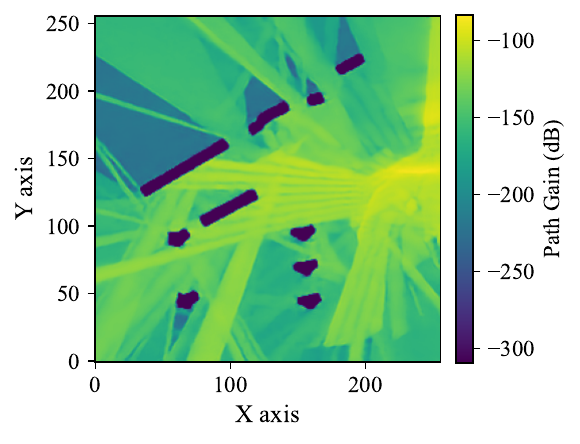}}
		\subfloat{\includegraphics[width=1.35in]{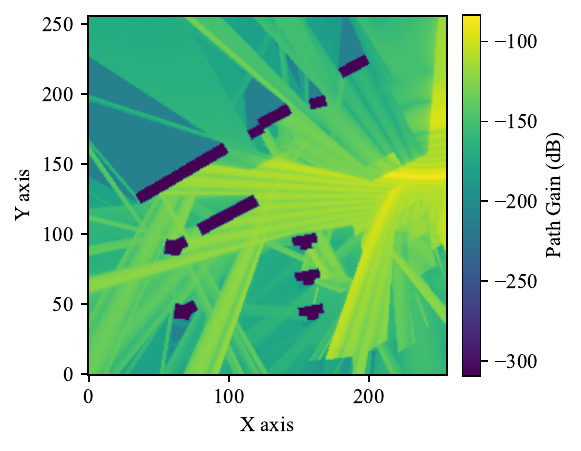}}
		\\
		\subfloat{\includegraphics[width=1.35in]{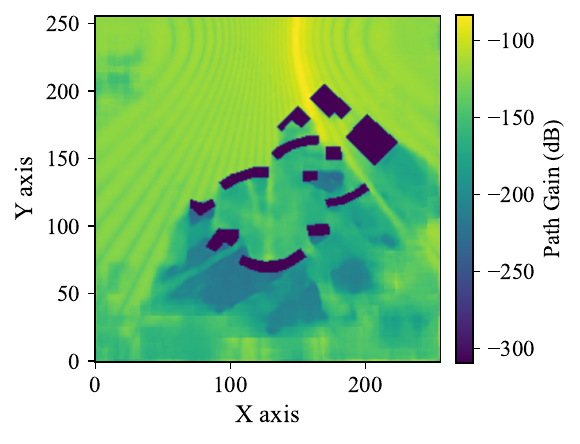}}
		\subfloat{\includegraphics[width=1.35in]{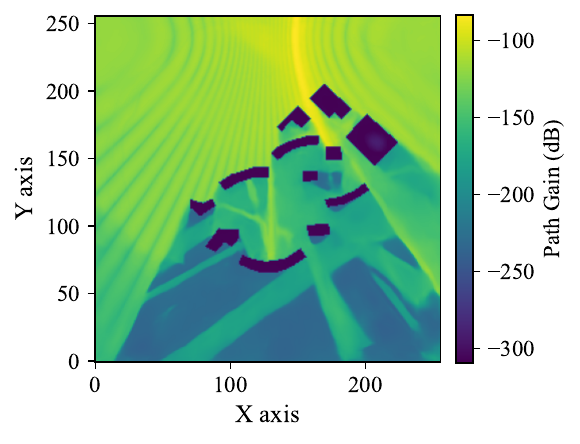}}
		\subfloat{\includegraphics[width=1.35in]{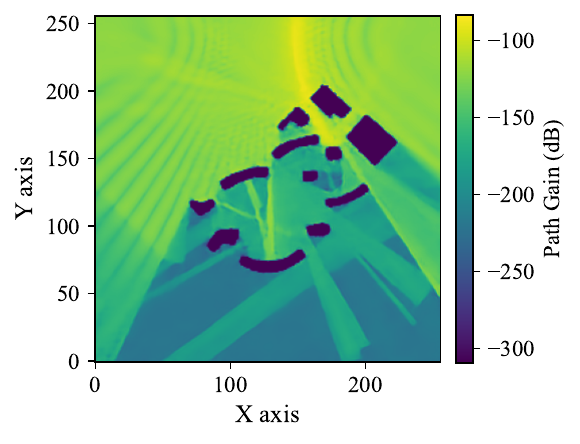}}
		\subfloat{\includegraphics[width=1.35in]{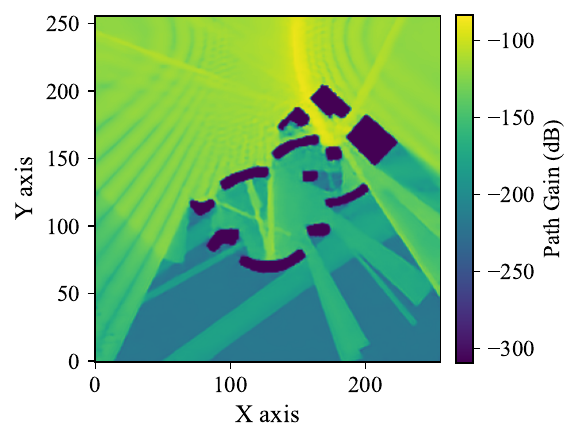}}
		\subfloat{\includegraphics[width=1.35in]{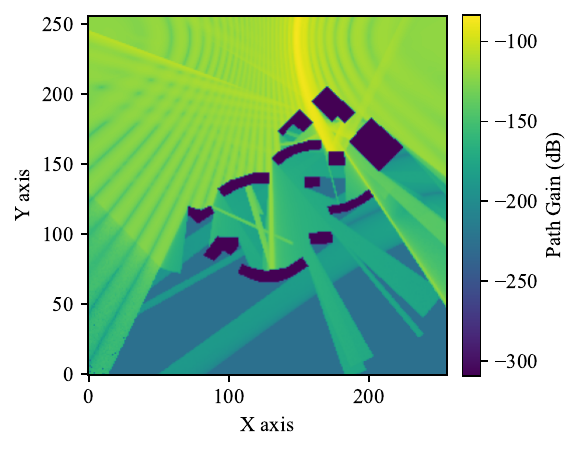}}
		\\
		\setcounter{subfigure}{0}
		\subfloat[PI-RadioUNet]{\includegraphics[width=1.35in]{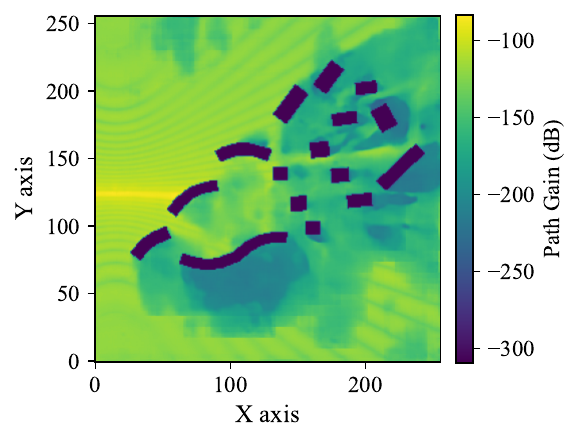}}
		\subfloat[PI-TransUNet]{\includegraphics[width=1.35in]{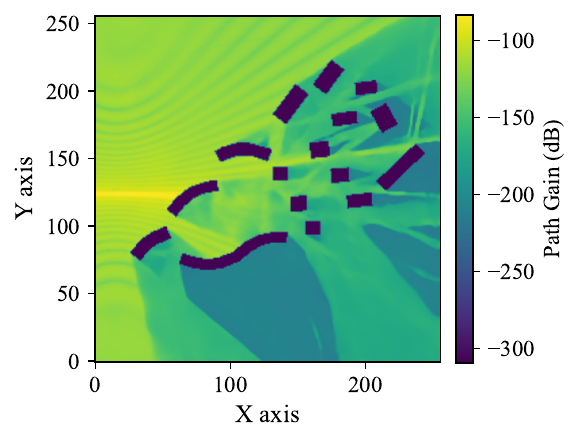}}
		\subfloat[PI-CKMDiff]{\includegraphics[width=1.35in]{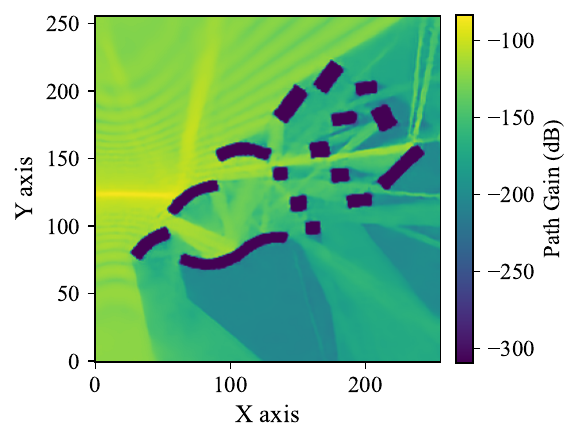}}
		\subfloat[BeamCKMDiff]{\includegraphics[width=1.35in]{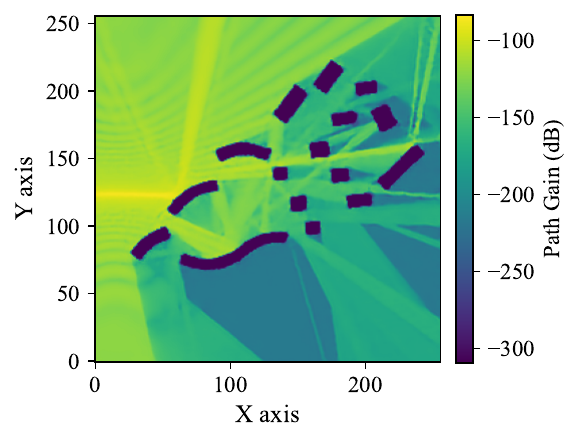}}
		\subfloat[Ground truth]{\includegraphics[width=1.35in]{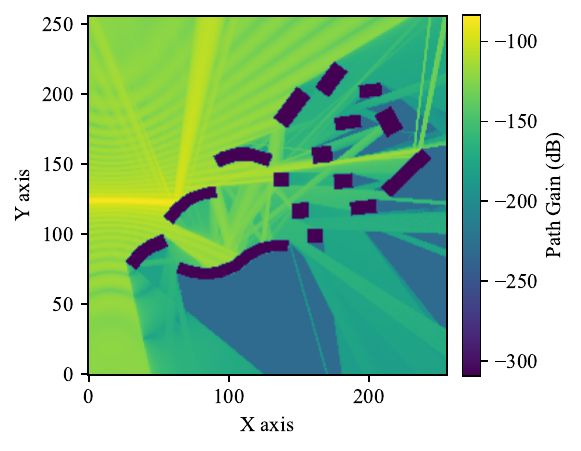}}
		\\
		\caption{Visual comparisons of the generated BeamCKMs. The results are generated across varying environmental topologies and continuous beamforming vectors via (a) PI-RadioUNet, (b) PI-TransUNet, (c) PI-CKMDiff, (d) BeamCKMDiff, with the visualization of (e) as ground truth.}
		%		\caption{The comparisons of constructed CKM in }
		\label{fig:visualization_umseemed_beam}
		\vspace*{-0.3cm}
	\end{figure*}

	\begin{figure}[!t]
		\centering
		\vspace{-0.3cm}
		\subfloat[]{\includegraphics[width=1.35in]{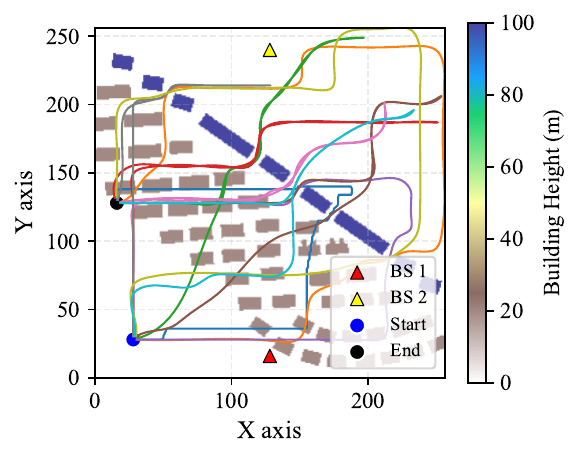}}
		\hspace{0.1cm}
		\subfloat[]{\includegraphics[width=1.35in]{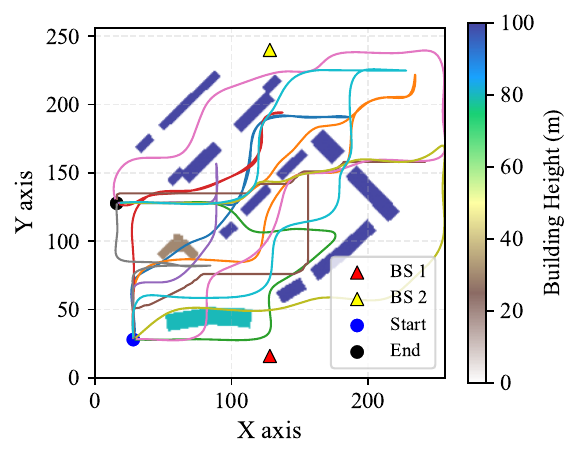}}
		\caption{{Visualization of trajectories of  (a) scenario \#$1$, (b) scenario \#$2$.}}
		\label{fig:trajectories_2_scenarios}
		\vspace*{-0.5cm}
	\end{figure}
	
	\begin{table}[!t]
		\renewcommand{\arraystretch}{1.15}
		\centering
		\caption{NMSE Performance Comparison}
		\label{tab:NMSE_performance_comparison}
		\setlength{\tabcolsep}{2mm}
		\begin{tabular}{l|c|c}
			\Xhline{1.1pt}
			\textbf{Methods} & \textbf{NMSE} (dB) & \textbf{Inference time} (s)\\
			\hline
			{PI-RadioUNet}        & $-17.56$ & $0.063$\\
			{PI-TransUNet}        & $-20.79$ & $0.065$\\
			{PI-CKMDiff} ($T_{\rm step} = 5$)  & $-21.26$ & $0.042$\\
			BeamCKMDiff ($T_{\rm step} = 2$) & $-22.66$ & $\mathbf{0.031}$\\
			BeamCKMDiff ($T_{\rm step} = 5$) & $-\mathbf{23.96}$ & $0.067$\\
			BeamCKMDiff ($T_{\rm step} = 10$) & $-23.79$ & $0.103$\\
			BeamCKMDiff ($T_{\rm step} = 50$) & $-23.75$ & $0.354$\\
			\Xhline{1.1pt}
		\end{tabular}
		\vspace*{-0.3cm}
	\end{table}

	\subsection{Performance of BeamCKM Construction}

	We compare BeamCKMDiff against three representative methods. {To ensure a fair comparison, we construct physics-informed (PI) baselines by integrating} the analytical LoS beam gain prior $\overline{\mathbf{\Psi}}_{\mathbf{w}_b}$ proposed in Section \ref{subsec:analytical_prior} into their input condition pipelines, {empowering models originally designed for single antennas} to handle continuous beam-aware CKM generation. The benchmarking methods are introduced below:
	\begin{enumerate}
		\item {{PI-RadioUNet} \cite{levie2021radiounet}:} A discriminative U-Net model that learns a deterministic mapping from environmental geometries to CKMs.
		\item {{PI-TransUNet} \cite{wang2025beamckm}:} A Transformer-UNet hybrid. We adopt its architecture and embed the continuous beamforming vector as a conditional token directly into the Transformer module.
		\item {{PI-CKMDiff} \cite{ZengYong_CKMDiff_arxiv}:} A U-Net-based diffusion model relying on standard convolutional denoising, lacking the adaLN mechanism for global conditioning.
	\end{enumerate}

	\begin{figure*}[!t]
		\centering
		\subfloat[]{\includegraphics[width=0.525\columnwidth]{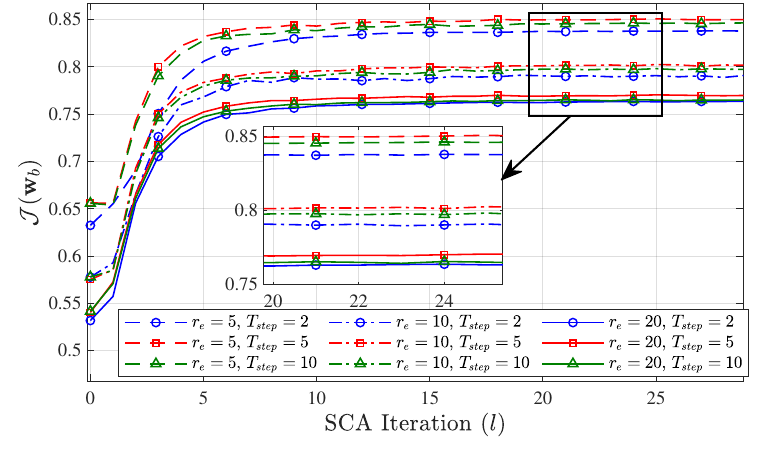}}
		\hspace{0.5cm}
		\subfloat[]{\includegraphics[width=0.525\columnwidth]{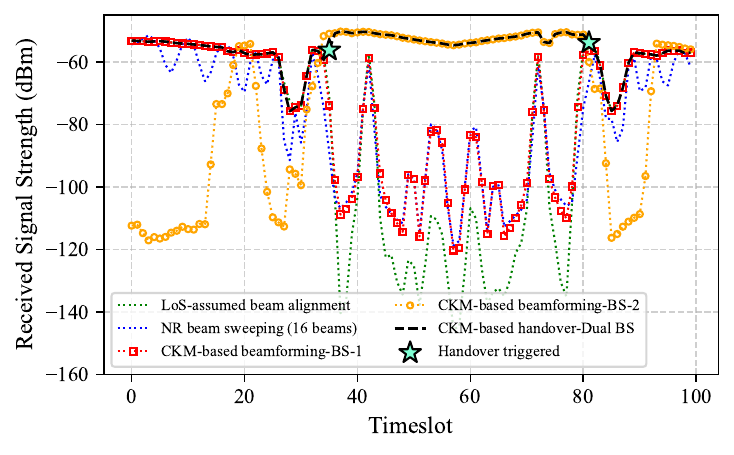}}
		\hspace{0.5cm}
		\subfloat[]{\includegraphics[width=0.525\columnwidth]{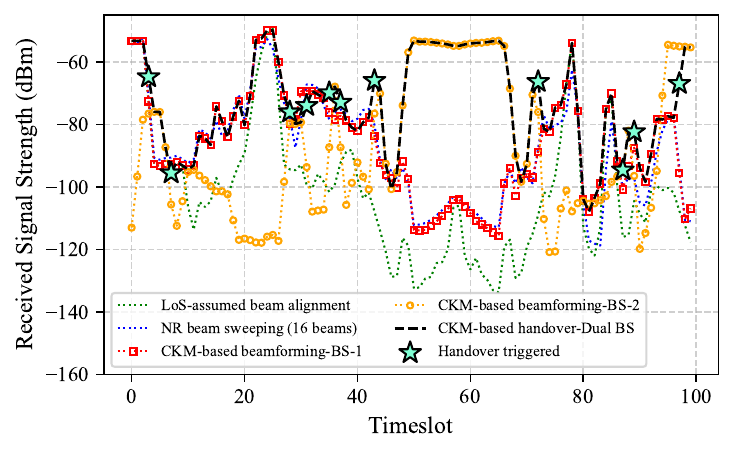}}
		\caption{{Performance of continuous CKM-based beamforming and proactive handover. Panel (a) reports convergence under different spatial-uncertainty radii and DDIM step counts; panels (b) and (c) report the RSS along two UAV trajectories for the compared beamforming schemes.}}
		\label{fig:SCA_Interation_and_one_path_results}
		\vspace*{-0.4cm}
	\end{figure*}

	\begin{figure*}[!t]
		\centering
		\hspace{-0.15cm}
		\subfloat[]{\includegraphics[width=1.35in]{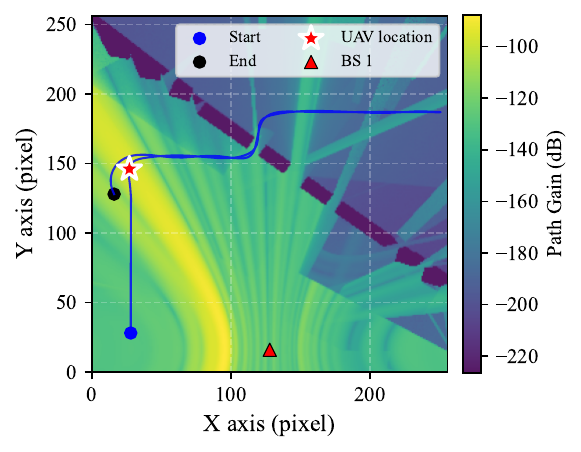}}
		\hspace{0.05cm}
		\subfloat[]{\includegraphics[width=1.65in]{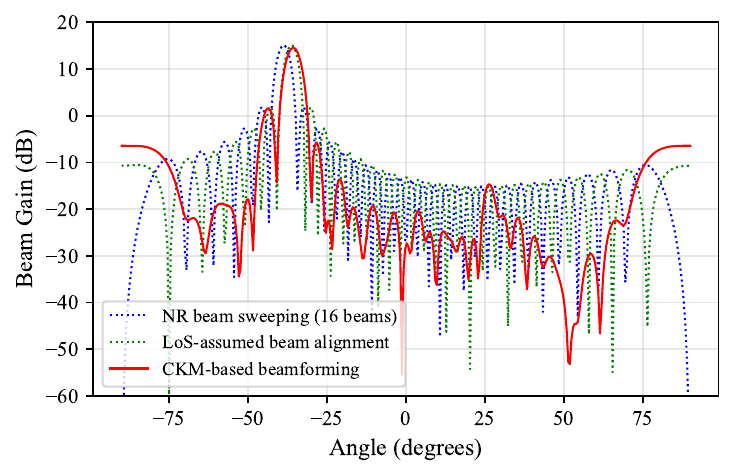}}
		\hspace{0.3cm}
		\subfloat[]{\includegraphics[width=1.35in]{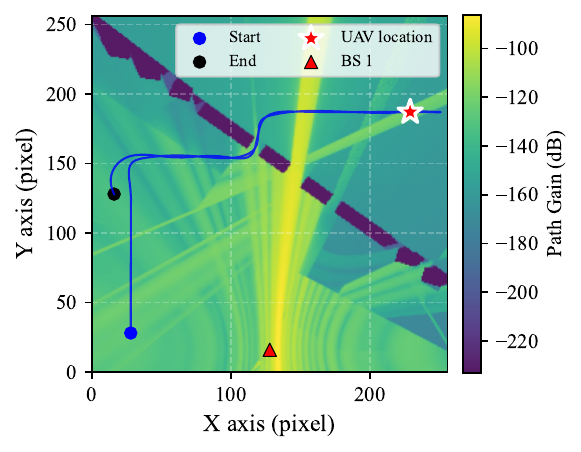}}
		\hspace{0.05cm}
		\subfloat[]{\includegraphics[width=1.65in]{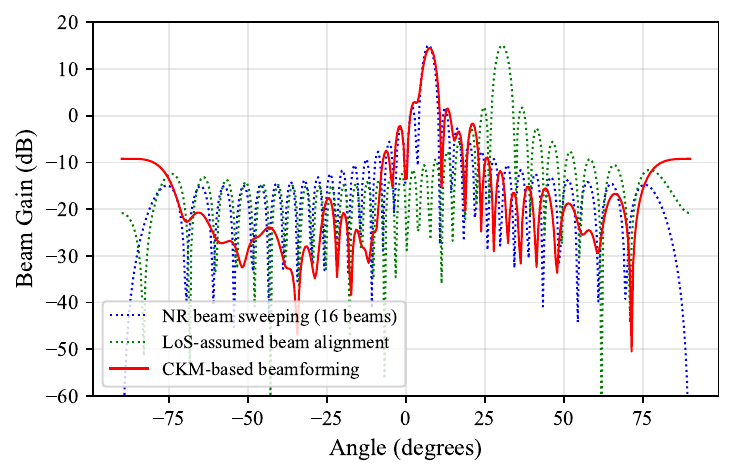}}
		\caption{Visualizations of the BeamCKMs and the corresponding angular beam gain patterns for UAV trajectory 3 in scenario $\#$1. The results depict the beamforming performance under (a)-(b) in the LoS scenario at step 18, and (c)-(d) in the NLoS blockage scenario at step 53.}
		\label{fig:result_trajetory_1}
		\vspace*{-0.4cm}
	\end{figure*}
	
	\begin{figure*}[!t]
		\centering
		\hspace{-0.15cm}
		\subfloat[]{\includegraphics[width=1.35in]{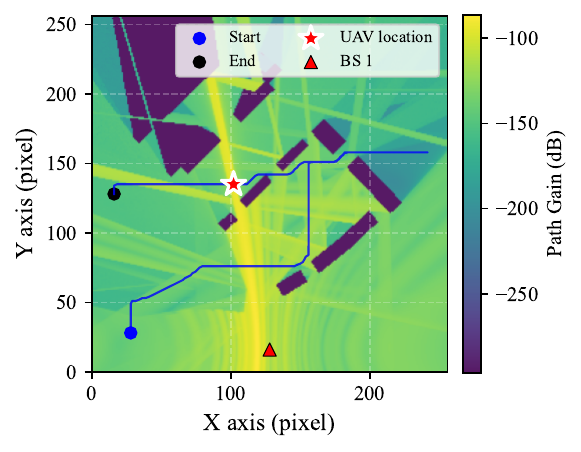}}
		\hspace{0.05cm}
		\subfloat[]{\includegraphics[width=1.65in]{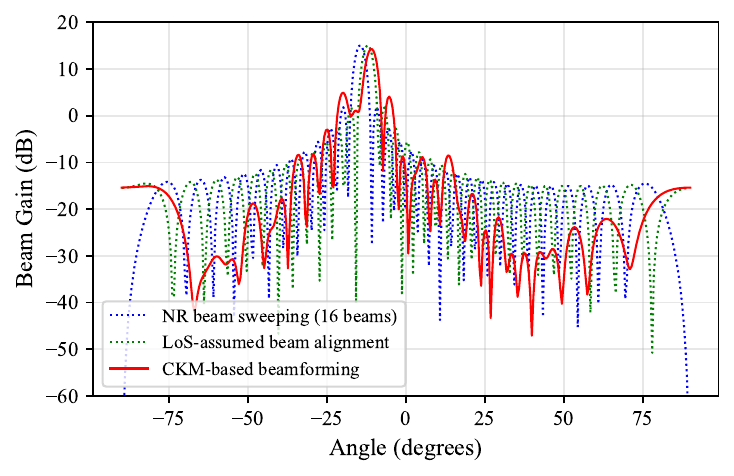}}
		\hspace{0.3cm}
		\subfloat[]{\includegraphics[width=1.35in]{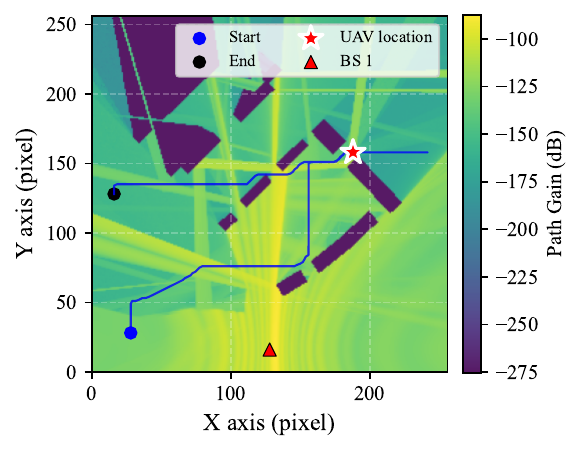}}
		\hspace{0.05cm}
		\subfloat[]{\includegraphics[width=1.65in]{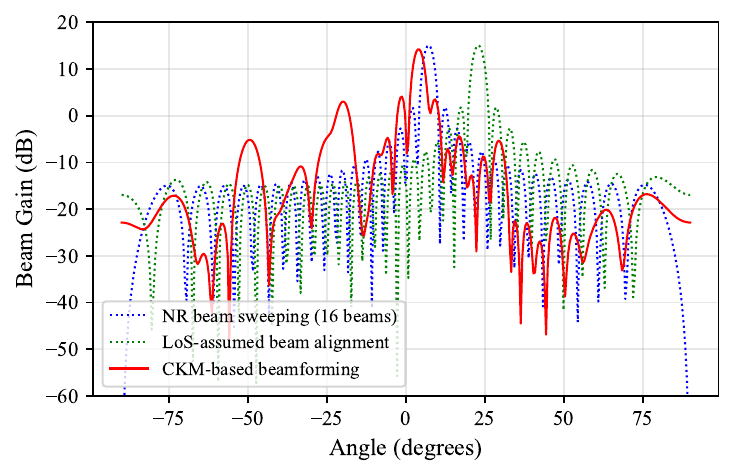}}
		\caption{Visualizations of the BeamCKMs and the corresponding angular beam gain patterns for UAV trajectory 5 in scenario $\#$2. The results depict the beamforming performance under (a)-(b) in the LoS scenario at step 84, and (c)-(d) in the NLoS blockage scenario at step 66.}
		\label{fig:result_trajetory_2}
		\vspace*{-0.35cm}
	\end{figure*}

	{We evaluate the constructed CKMs using normalized mean-square error (NMSE) computed in the linear power domain:
	\begin{equation}
		\label{nmse_define}
		{\rm NMSE (dB)} = 10 \log_{10} \left( \frac{\|\widehat{\mathbf{\Psi}}_{\rm lin} - \mathbf{\Psi}_{\rm lin}\|_F^2}{\|\mathbf{\Psi}_{\rm lin}\|_F^2} \right),
	\end{equation}
		where $\mathbf{\Psi}_{\rm lin}$ and $\widehat{\mathbf{\Psi}}_{\rm lin}$ are the ground-truth and predicted power-gain maps, respectively, and $\|\cdot\|_F$ is the Frobenius norm. The logarithm is applied only to report the final ratio in decibels.}

	\myreftable{tab:NMSE_performance_comparison} evaluates the generation accuracy, inference latency, and the impact of the sampling steps $T_{\text{step}}$ during the DDIM reverse process. For BeamCKMDiff, $T_{\text{step}} = 5$ provides the optimal trade-off, achieving an NMSE of $-23.96$ dB with a latency of $0.067$ s. Increasing $T_{\text{step}}$ further yields negligible accuracy gains while severely degrading latency. Under this configuration, BeamCKMDiff outperforms all baselines. {Although the physical prior significantly boosts PI-CKMDiff to an NMSE of $-21.26$ dB, BeamCKMDiff further improves the accuracy by $2.7$ dB.} Compared to {PI-RadioUNet} ($-17.56$ dB) and {PI-TransUNet} ($-20.79$ dB), it demonstrates superior capability in mapping continuous beamforming vectors to spatial gains. {Furthermore, its inference time is comparable to baselines, ensuring practical BeamCKM construction.}

	%	\myreffig{fig:visualization_umseemed_beam} visually compares the generated BeamCKMs under the optimal setting of $T_{\text{step}} = 5$. While the analytical prior helps all methods estimate the main lobe direction, CNN-based baselines rely on local feature fusion. Thus, under severe blockages (NLoS), they fail to infer non-local scattering and depict only generic shadowing. In contrast, BeamCKMDiff leverages the global attention of DiT and injects the beamforming vector via the adaLN mechanism to precisely reconstruct spatial propagation characteristics, such as reflections and diffractions.

	\myreffig{fig:visualization_umseemed_beam} visually compares the generated BeamCKMs. Since quantitative performance is evaluated across various DDIM steps, the visual comparisons are consistently presented using the optimal setting of $T_{\text{step}} = 5$. While the integrated analytical prior helps all methods estimate the main lobe direction, baselines rely primarily on local convolutional fusion. Thus, when the main lobe directly hits a blockage, they fail to infer non-local scattering, causing them to suffer from beam misalignment and depict only generic shadowing. In contrast, BeamCKMDiff leverages the global attention of DiT and injects the continuous beamforming vector directly as a global token via the adaLN mechanism. Furthermore, unlike PI-CKMDiff, which simply concatenates the beam condition as a spatial input and thus hinders gradient backpropagation, our adaLN-based beam conditioning ensures smooth gradient flow. By overcoming the vanishing gradient issue of standard convnets, BeamCKMDiff becomes fully differentiable, enabling the analytical beamforming optimization discussed in the next subsection.

	\vspace*{-0.1cm}
	\subsection{Performance of CKM-based Beamforming and Handover}

	The proposed CKM-based beamforming and handover methods are benchmarked against two traditional methods:
	\begin{enumerate}
		\item LoS-assumed beam alignment \cite{Beamforming_Mag}, which aligns beams geometrically based on purely LoS propagation, ignoring environmental blockages.
		\item NR beam sweeping \cite{BeamSweeping_CM}, which sweeps across a predefined discrete fourier transform (DFT) codebook of 16 directional beams within the $[0, \pi]$ range.
	\end{enumerate}

	To evaluate the statistical robustness of the proposed framework, Monte Carlo simulations are conducted using 10 random UAV trajectories across two distinct complex urban scenarios, as depicted in \myreffig{fig:trajectories_2_scenarios}. {In each scenario, the UAV navigates from a fixed start point to a fixed end point over a flight period of $100$ s, with a randomly generated trajectory.} Two BSs are deployed to serve the moving UAV. For the dynamic trajectory evaluation, the proposed{proactive} dual-BS CKM-based handover is compared against single-BS CKM-Beamforming. {Note that all reported results below are evaluated over the exact ground-truth channels rather than the predicted BeamCKM.}

	In \myreffig{fig:SCA_Interation_and_one_path_results}(a), we evaluate the convergence of the optimized target function $\mathcal{J}(\mathbf{w}_b)$ under varying spatial constraints $r_e \in \{5, 10, 20\}$ and DDIM steps $T_{\rm step} \in \{2, 5, 10\}$. Consistent with the previous quantitative analysis in \myreftable{tab:NMSE_performance_comparison}, $T_{\rm step}=5$ achieves the highest objective value. Increasing the steps further degrades optimization performance, likely due to the accumulated gradient variance through the unrolled reverse diffusion process. Additionally, a tighter spatial constraint, corresponding to higher UAV location precision, inherently yields a higher concentrated signal gain. The SCA algorithm demonstrates rapid convergence, stabilizing within 15 iterations across all configurations. Consequently, we fix $r_e=5$ and $T_{\rm step}=5$ for all subsequent dynamic evaluations.

	\begin{figure*}[!t]
		\centering
		\subfloat[]{\includegraphics[width=0.525\columnwidth]{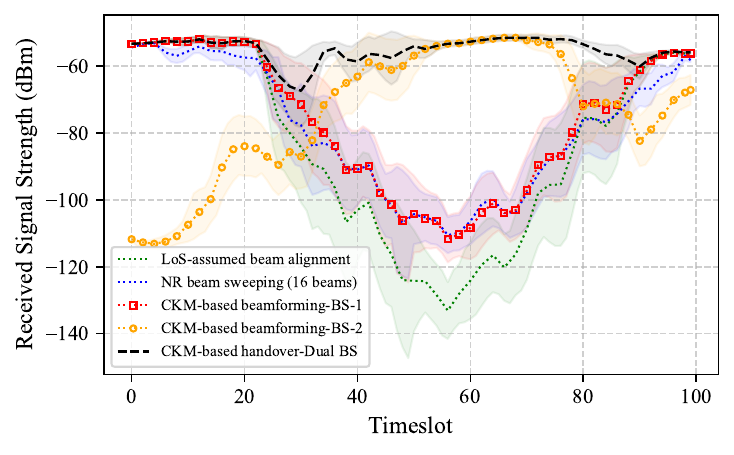}}
		\hspace{0.5cm}
		\subfloat[]{\includegraphics[width=0.525\columnwidth]{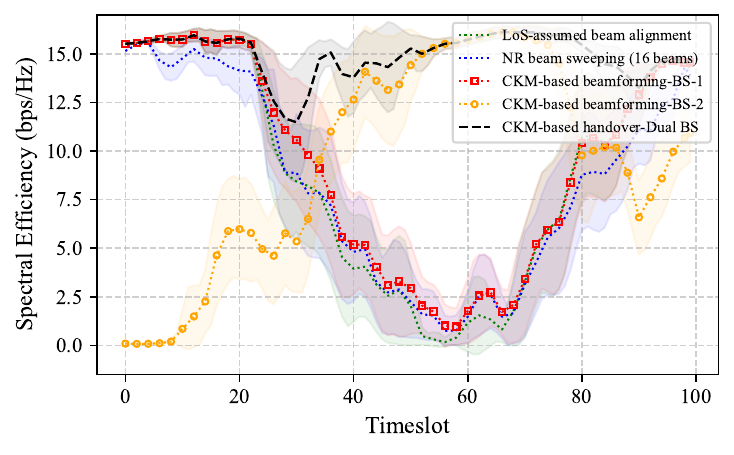}}
		\hspace{0.5cm}
		\subfloat[]{\includegraphics[width=0.525\columnwidth]{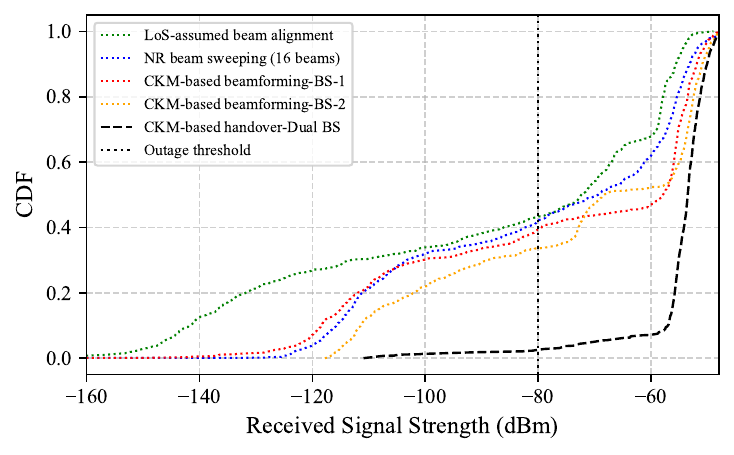}}
		\caption{(a) Received signal strength
			(b) spectral efficiency varying over time
			{(c) CDF of received signal strength for trajectories in scenario $\#$1.}}
		\label{fig:results_10_trajs_sce_1}
		\vspace*{-0.35cm}
	\end{figure*}
	
	\begin{figure*}[!t]
		\centering
		\subfloat[]{\includegraphics[width=0.525\columnwidth]{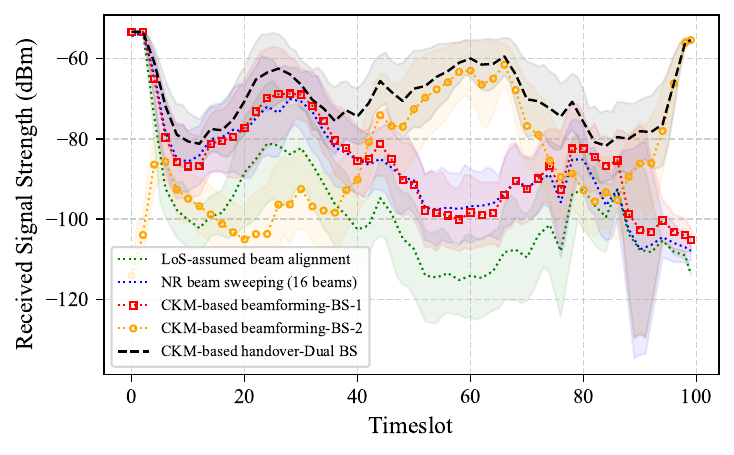}}
		\hspace{0.5cm}
		\subfloat[]{\includegraphics[width=0.525\columnwidth]{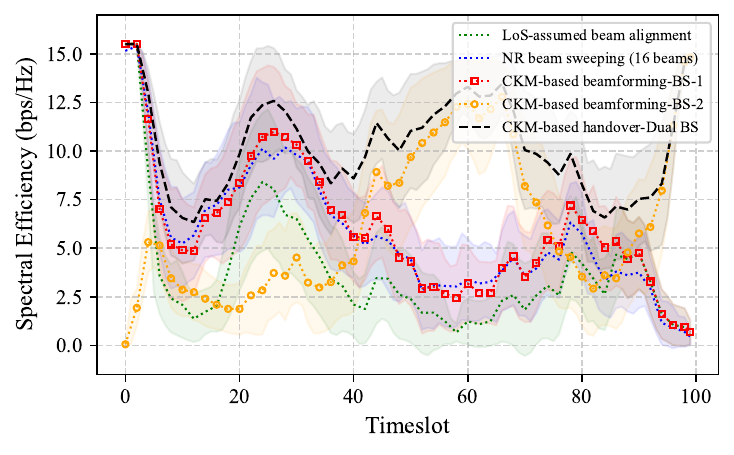}}
		\hspace{0.5cm}
		\subfloat[]{\includegraphics[width=0.525\columnwidth]{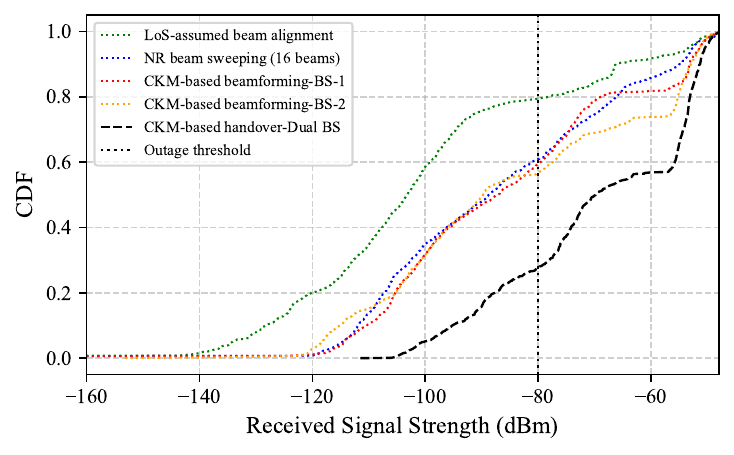}}
		\caption{(a) Received signal strength
			(b) spectral efficiency varying over time
			{(c) CDF of received signal strength for trajectories in scenario $\#$2.}}
		\label{fig:results_10_trajs_sce_2}
		\vspace*{-0.35cm}
	\end{figure*}
	
	\myreffig{fig:SCA_Interation_and_one_path_results}(b) and (c) illustrate the received signal strength over timeslots for two distinct UAV trajectories. The proposed continuous CKM-Beamforming consistently outperforms the baselines. Specifically, NR beam sweeping suffers from quantization errors due to its discrete angular codebook, while the LoS-Assumption scheme experiences severe signal drops when encountering environmental NLoS conditions.
	Regarding the{proactive} handover, the dual-BS CKM-based handover mechanism effectively maintains optimal link quality. In \myreffig{fig:SCA_Interation_and_one_path_results}(b), the UAV traverses the coverage areas with minimal blockage, executing only two handovers to switch from BS 1 to BS 2 and back. Conversely, \myreffig{fig:SCA_Interation_and_one_path_results}(c) depicts a more complex scattering environment that necessitates more frequent BS switching to mitigate sudden signal drops. The dual-BS coordination successfully prevents deep fades, confirming its capability to ensure link stability in highly obstructed aerial environments.

	To further elucidate the performance gains, \myreffig{fig:result_trajetory_1} and \myreffig{fig:result_trajetory_2} visualize the spatial BeamCKMs and the corresponding angular beam gain patterns at specific LoS and NLoS beamforming steps.
	In the LoS scenarios (\myreffig{fig:result_trajetory_1}(a)-(b) and \myreffig{fig:result_trajetory_2}(a)-(b)), both the LoS-Assumption scheme and the proposed CKM-Beamforming effectively align the main lobe with the UAV location. In contrast, the discrete NR Beam Sweeping scheme exhibits noticeable quantization errors, failing to achieve perfect alignment due to its limited codebook resolution.
	The distinct advantage of the proposed framework is prominently demonstrated in the NLoS scenarios (\myreffig{fig:result_trajetory_1}(c)-(d) and \myreffig{fig:result_trajetory_2}(c)-(d)). When the direct path is severely obstructed by buildings, the LoS-Assumption method blindly steers the main beam directly into the physical obstacles, leading to severe signal degradation. Conversely, CKM-Beamforming leverages the learned spatial environmental awareness to autonomously adapt its continuous beamforming vector. It effectively synthesizes deflected or multi-lobe beam patterns to exploit available reflections and diffractions, thereby bypassing the blockages and maintaining a robust communication link.

	\myreffig{fig:results_10_trajs_sce_1} and \myreffig{fig:results_10_trajs_sce_2} illustrate the mean and variance of the received signal strength and spectral efficiency over time for the two scenarios. Statistically, the single-BS CKM-Beamforming consistently outperforms both the NR Beam Sweeping and LoS-Assumption schemes, effectively mitigating NLoS deep fades. Furthermore, the dual-BS CKM-based handover dynamically selects the superior propagation path, seamlessly tracking the upper performance envelope to maximize spectral efficiency and ensure robust link stability throughout the flight.
	The cumulative distribution functions (CDFs) in \myreffig{fig:results_10_trajs_sce_1}(c) and \myreffig{fig:results_10_trajs_sce_2}(c) further validate these gains.{At the $-80$ dBm outage threshold, the conventional baselines exhibit substantial outage probabilities because they cannot adapt to complex multipath propagation. The proposed dual-BS handover produces no observed outage samples and provides the strongest empirical coverage over the evaluated trajectories.}

\begin{figure}[!t]
		\centering
		\includegraphics[width=0.55\linewidth]{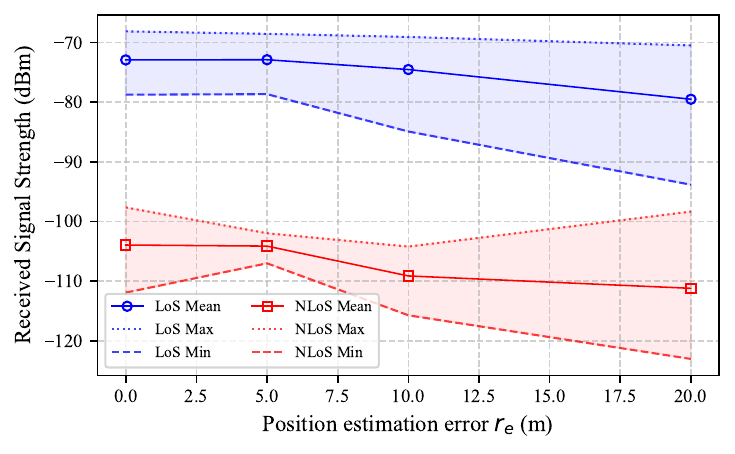}
		\caption{{Achievable RSS at 1,000 test locations under position-error bounds $r_{\rm e}\in\{0,5,10,20\}$ m. Each beam is evaluated at the ground-truth UAV coordinate using Sionna ray tracing.}}
		\label{fig:pos_err_robust}
		\vspace*{-0.15cm}
	\end{figure}

\begin{figure}[!t]
		\centering
		\includegraphics[width=0.55\linewidth]{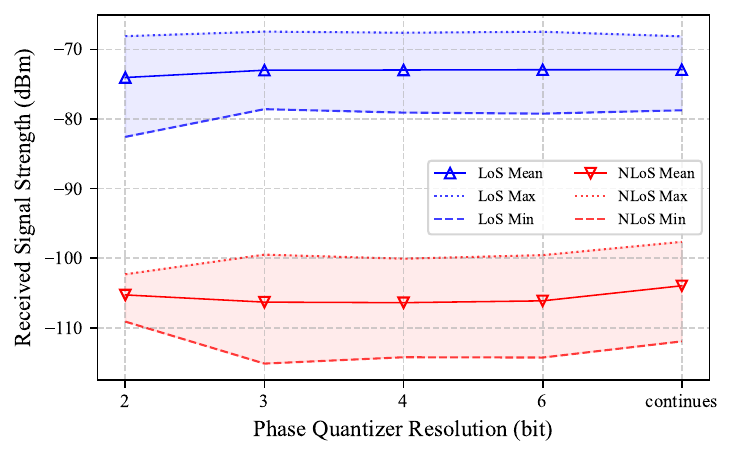}
		\caption{{Achievable RSS at 1,000 test locations under continuous beamforming and $b$-bit phase quantization for $b\in\{2,3,4,6\}$, evaluated using Sionna ray tracing.}}
		\label{fig:quant_robust}
		\vspace*{-0.4cm}
	\end{figure}

	{We further evaluate physical beamforming gains at 1,000 LoS and NLoS test locations using Sionna ray tracing. For localization robustness, beams are optimized with error bounds $r_{\rm e}\in\{0,5,10,20\}$ m and evaluated at the ground-truth coordinates. Fig.~\ref{fig:pos_err_robust} shows a degradation below $2.5$ dB for $r_{\rm e}\leq10$ m relative to error-free positioning. {To evaluate the impact of phase quantization}, the optimized beam $\mathbf{w}^*$ is projected onto a constant-envelope $b$-bit phase-shifter array according to
	\begin{equation}
		[\widehat{\mathbf{w}}]_m=\frac{1}{\sqrt{N_t}}\exp\!\left(j\Delta\theta\left\lfloor\frac{\angle[\mathbf{w}^*]_m}{\Delta\theta}\right\rceil\right),\,\, \Delta\theta=\frac{2\pi}{2^b},
	\end{equation}
	where $\lfloor\cdot\rceil$ denotes nearest-integer rounding. {Compared to the position error, Fig.~\ref{fig:quant_robust} shows a more moderate degradation from phase quantization, where the continuous beam yields the highest RSS and low-resolution quantization maintains robust performance in both LoS and NLoS conditions.}}

	Beyond performance enhancements, BeamCKMDiff also shifts resource utilization from the air interface to the computational domain. It has the potential to reduce reliance on exhaustive NR sweeping by predicting the optimal continuous beam from CKM and UAV coordinates, thereby helping the serving BS bypass physical sweeping. This paradigm can also extend to{proactive} handover, where virtual link evaluations may reduce the need for candidate BSs to broadcast physical scanning pilots.{Edge offloading removes neural inference and beam optimization from the UAV, preserving its battery for propulsion and mission payload. The resulting design replaces repeated air-interface pilot transmission with grid-powered edge computation. Edge caching and UAV energy-management methods can further improve system-level efficiency \cite{uavcaching2025}.}

	%	\vspace*{-0.2cm}
	\section{Conclusion}
	\label{sec:conclusion}
	In this paper, we proposed BeamCKMDiff, a physics-informed diffusion framework designed to overcome the prohibitive pilot training overhead in dynamic UAV communications. By integrating an analytical spatial prior with a DiT backbone via an adaLN mechanism, the framework accurately maps continuous beamforming vectors to high-fidelity channel gain distributions. Leveraging the end-to-end differentiability of the deterministic reverse diffusion process, we formulated a CKM-triggered beamforming and proactive dual-BS handover framework. This enables continuous beam optimization via SCA without executing exhaustive physical pilot sweeps over the air interface. Numerical evaluations verified that BeamCKMDiff achieves NMSE of $-23.96$ dB with sub-$100$ ms latency. Furthermore, the predictive dual-BS handover effectively eliminates deep fades in NLoS environments, minimizing outage probabilities and maximizing effective SE. Future work will focus on extending this proactive spatial inference paradigm to multi-UAV cooperative networks and exploring its potential in aerial swarm communications.{Future directions also include routing-aware coordination strategies \cite{etar2015} for dynamic UAV networks under mobility constraints, altitude-varying BeamCKM construction, and the robustness of the LoS beam prior under extreme weather-induced fading and imperfect environmental topology.}

	\bibliographystyle{IEEEtran}
	\bibliography{IEEEabrv, Ref_revised}

\end{document}